\documentclass[tradiabstract]{aa} 

\usepackage{graphicx}
\usepackage{txfonts}
\usepackage{color}

\usepackage{amsmath,amsfonts,amssymb}
\usepackage{fixltx2e}
\usepackage{caption,subcaption}
\usepackage[breaklinks, colorlinks, citecolor=blue]{hyperref}
\usepackage{natbib}
\usepackage{tabularx}
\usepackage[export]{adjustbox}

\bibpunct{(}{)}{;}{a}{}{,}
\newcommand{\Planck}{{\em Planck}}

\newcommand{\Rfive}{R_{500}}
\newcommand{\thetas}{\theta_{\rm s}}

\newcommand{\yo}{y_{\rm o}}     
\newcommand{\estyo}{\hat{y}_{\rm o}}
\newcommand{\fnu}{f_\nu}                
\newcommand{\Tth}{T_{\thetas}}
\newcommand{\tth}{t_{\thetas}}

\newcommand{\Psit}{\Psi_{\thetas}}
\newcommand{\sigt}{\sigma_{\thetas}}

\newcommand{\Om}{\Omega_\textrm{m}}
\newcommand{\sig}{\sigma_\textrm{8}}

\begin{document}

   \title{PSZSPT: A joint \Planck\ and SPT-SZ cluster catalog}
     \author{J.-B. Melin
          \inst{1}
          \and J. G. Bartlett
           \inst{2,3}
           \and P. Tarr{\'i}o
           \inst{4,1}
           \and
           G. W. Pratt
           \inst{4,1}
               }

   \institute{IRFU, CEA, Universit{\'e} Paris-Saclay, F-91191 Gif-sur-Yvette, France\\
                 \email{jean-baptiste.melin@cea.fr}
         \and
                 Universit\'e de Paris, CNRS, Astroparticule et Cosmologie,  F-75006 Paris, France
            \and
                  Jet Propulsion Laboratory, California Institute of Technology, 4800 Oak Grove Drive, Pasadena, California, USA
            \and AIM, CEA, CNRS, Universit{\'e} Paris-Saclay, Universit{\'e} Paris Diderot, Sorbonne Paris Cit{\'e}, F-91191 Gif-sur-Yvette, France
             }

   \date{Received ...; accepted ...}

% \abstract{}{}{}{}{} 
% 5 {} token are mandatory
 
 \abstract
{We present the first cluster catalog extracted from combined space-based (\Planck) and ground-based (South Pole Telescope; SPT-SZ) millimeter data. We developed and applied a matched multi-filter (MMF) capable of dealing with the different transfer functions and resolutions of the two datasets. We verified that it produces results consistent with publications from \Planck\ and SPT collaborations when applied on the datasets individually. We also verified that \Planck\ and SPT-SZ cluster fluxes are consistent with each other.  When applied blindly to the combined dataset, the MMF generated a catalog of 419 detections ($S/N>5$), of which 323 are already part of the SPT-SZ or PSZ2 catalogs; 54 are new SZ detections, which have been identified in other catalogs or surveys; and 42 are new unidentified candidates. The MMF takes advantage of the complementarity of the two datasets, \Planck\ being particularly useful for detecting clusters at a low redshift ($z<0.3$), while SPT is efficient at finding higher redshift ($z>0.3$) sources. This work represents a proof of concept that blind cluster extraction can be performed on combined, inhomogeneous millimeter datasets acquired from space and ground. This result is of prime importance for planned ground-based cosmic microwave background (CMB) experiments (e.g., Simons Observatory, CMB-S4) and envisaged CMB space missions (e.g., {\em PICO}, {\em Backlight}) that will detect hundreds of thousands of clusters in the low mass regime ($M_{500} \leqslant 10^{14} M_\odot$), for which the various sources of intra-cluster emission (gas, dust, synchrotron) will be of the same order of magnitude and hence require broad ground and space frequency coverage with a comparable spatial resolution for adequate separation.}


   \maketitle
%
%-------------------------------------------------------------------

\section{Introduction}

Galaxy clusters constitute unique objects to study structure formation in the Universe. Lying at the nodes of the cosmic web, their distribution in mass and redshift is sensitive to cosmological parameters~\citep[e.g.,][and references therein]{allen2011}. They also represent ideal laboratories to understand galaxy formation and evolution~\citep[e.g.,][and references therein]{voit2005}.
Progressing in both cosmology and astrophysics with galaxy clusters requires advances in two directions: increasing the number of known clusters and better understanding their physics. The multi-frequency view is now mandatory to achieve these goals. 

Clusters are detected in the optical via their member galaxies, in the X-ray via Bremsstrahlung emission of the embedded hot ionized gas and, more recently, via the Sunyaev-Zeldovich~\citep[SZ,][]{sunyaev1970,sunyaev1972} effect.  Progress in detecting numerous new clusters has been made in recent years with the advent of SZ surveys~\citep{staniszewski2009, menanteau2010, ESZ}. The current galaxy cluster catalogs are most often extracted from single experiment data. This is the case for the all-sky Abell~\citep{abell1989}, ROSAT~\citep{noras,reflex}, and \Planck~\citep{ESZ,PSZ1,PSZ2} catalogs. 

In order to extract clusters with a low signal-to-noise threshold, hybrid methods have been developed recently to clean the single experiment catalogs from spurious detections. These methods have been successfully applied to X-ray (ROSAT) and SZ (\Planck) detections using optical (BOSS, DES, WISE, SDSS) data~\citep[e.g.,][]{burenin2017,finogenov2020,klein2019}. Another approach to extract new clusters is to use data from different frequency bands jointly. The approach was initially proposed by~\cite{maturi2007}, but it is difficult to implement in practice because the signals from clusters originate from different physical processes in different frequency bands. It has, nevertheless, been successfully implemented recently by \cite{tarrio2016, tarrio2018, tarrio2019} for ROSAT and \Planck\ data. 

A final avenue for extracting low signal-to-noise clusters is to combine different observations in the same frequency band. \cite{aghanim2019}, \cite{madha2020}, and \cite{naess2020} followed this path and produced, for the first time, combined SZ maps from \Planck\ and ACT data, \Planck\ and ACTPol data, as well as \Planck\ and ACT, ACTPol, and AdvACT data. These works showed, for the first time, the significant gain for cluster science when combining space- and ground-based data in the ACT footprint. \cite{aghanim2019} also extracted clusters using a match multi-filter (MMF), but they did not publish the associated catalog. 

Here, we focus on the \Planck\ and SPT-SZ datasets. We propose a practical implementation of a blind MMF extraction algorithm working on space- and ground-based data jointly, and we publish the associated catalog: PSZSPT. Cross-matches of the PSZSPT detections and external catalogs are included. We did not use the \Planck\ \& SPT-SZ combined maps proposed by~\cite{chown2018} to produce the PSZSPT catalog because they are not optimized for cluster extraction. We used the SPT maps  instead, which are provided in that publication, and we thoroughly tested them against quantities published with the official SPT-SZ catalogs; only then did we combine them with the public \Planck\ maps.

The study presented here is expected to be important for the forthcoming space- and ground-based experiments. The future CMB experiments will detect low mass clusters ($M_{500}\leqslant10^{14} M_\odot$) for which the SZ signal is expected to be of the same order as the other sources of emission from clusters, in particular the radio and infrared emission of galaxies~\citep[e.g.,][]{melin2018}. Combining ground-based ($\nu<300 \, {\rm GHz}$) and  space-based ($\nu>300 \, {\rm GHz}$) experiments would help disentangle the various sources of emission [e.g., Simons Observatory~\citep{simonsobservatory} \& \Planck\ or CMB-S4~\citep{cmbs4} \& {\em PICO}~\citep{hanany2019} or {\em Backlight}~\citep{basu2019,delabrouille2019}].

We first present the two datasets used in our analysis in Sect.~\ref{sec:datasets}. We then recall the characteristics of the MMF in Sect.~\ref{sec:mmf}. We apply the MMF on the SPT-SZ and \Planck\ data independently to test the consistency of our results with results published by the two collaborations in Sect.~\ref{sec:public_consistency}. The in-depth work related to Sect.~\ref{sec:public_consistency} is presented in Appendix~\ref{sec:filterfunc}, \ref{sec:sptszphot}, and \ref{sec:plckphot}. In Sect.~\ref{sec:sptplanck_consistency}, we test the consistency between the SPT-SZ and the \Planck\ datasets. We detail the construction of the PSZSPT catalog in Sect.~\ref{sec:pszspt} and how we characterize it. We provide a comparison between recovered and published masses in Appendix~\ref{sec:masscomp}. We give the names of the SPT and PSZ2 clusters missed in the PSZSPT catalog in Appendix~\ref{sec:missed_names}. The description of the PSZSPT catalog is given in Appendix~\ref{sec:description}. Finally, we summarize and discuss our findings, and look to future work in Sect.~\ref{sec:sum}.

\section{Data sets}
\label{sec:datasets}

\subsection{SPT-SZ}

We use the SPT-SZ public maps\footnote{\url{https://lambda.gsfc.nasa.gov/product/spt/spt_prod_table.cfm}} "SPT Only Data maps" at 95, 150, and 220~GHz. The maps provide a resolution of 1.75~arcmin (full width at half maximum, FWHM), which is slightly degraded with respect to the native resolution of 1.6, 1.1, and 1.0~arcmin at 95, 150, and 220~GHz, respectively~\citep{bleem2015}. The other key ingredients for the analysis described in this paper are the filter transfer functions, "SPT Filter Transfer Function" at each frequency and the boundary and point source mask "Mask". All the products are provided at Healpix $N_{side}=8192$, which corresponds to a pixel size of about 0.43~arcmin. The frequency responses are not provided in electronic format in the archive. We retrieved them from Fig.~10 of~\cite{chown2018} (long dashed lines in the three panels) using the WebPlotDigitizer\footnote{\url{https://automeris.io/WebPlotDigitizer/}}. For additional details about the SPT-SZ public data, we refer the reader to the LAMBDA archive webpage given in footnote 1 and to~\cite{chown2018}.

\subsection{\Planck}

We used the public \Planck\ maps of the High Frequency Instrument covering the six frequencies 100, 143, 217, 353, 545, and 857~GHz. The maps are provided in Galactic coordinates at Healpix resolution $N_{side}=2048$, corresponding to a pixel size of about 1.72~arcmin. We upgraded the maps to $N_{side}=8192$ by zero padding in  harmonic space, and we changed the coordinate system to equatorial to match the SPT-SZ public data. For the analysis, we assume the \Planck\ beams are Gaussian with an FWHM of 9.659, 7.220, 4.900, 4.916, 4.675, and 4.216~arcmin at 100, 143, 217, 353, 545, and 857~GHz, respectively, as in~\cite{PSZ2}. The frequency responses are also based on the same reference. We converted the maps to $\mu {\rm K}$, that is to say the units of the SPT-SZ maps.

\section{Matched multi-filters}
\label{sec:mmf}

We modified the MMFs, MMF3~\citep{melin2006,melin2012}, initially based on~\citet{herranz2002}. The MMF3 algorithm was developed to extract clusters from the \Planck\ maps for the three data releases~\citep{ESZ, PSZ1, PSZ2}. It works on \Planck\ data with $10 \times 10 \deg$ tangential maps. We wanted to keep the same \Planck\ size for the tangential maps to ease the component separation on a large scale and the SPT-SZ pixel size to conserve the information at a small scale.  We therefore divided the six \Planck\ and three SPT-SZ frequency maps covering the SPT footprint into $10 \times 10 \deg$ tangential maps and kept the 0.43~arcmin pixels corresponding to $N_{side}=8192$. We filtered the resulting maps of $1400 \times 1400$ pixels with MMF3. These maps contain $4^2=16$ times more pixels than for the standard \Planck\ analysis and nine frequency maps instead of six, leading to a computationally heavier analysis.

We write the nine tangential maps as $\vec{m}(\vec{x})$ and decomposed them as the cluster component $\yo\vec{\tth}(\vec{x})$ and the noise $\vec{n}(\vec{x})$, corresponding to both instrumental noise and astrophysical components other than the cluster:
\begin{equation}
\label{eq:datamodel}
\vec{m}(\vec{x}) =  \yo\vec{\tth}(\vec{x}) + \vec{n}(\vec{x}),
\end{equation}
where $\yo$ is the central Compton-$y$ parameter, $\vec{\tth}(\vec{x})$ is the vector whose $i^{th}$ component is \mbox{$\fnu(\nu_i) [b_i\ast \Tth](\vec{x})$}, the tSZ template $\Tth$ convolved by $b_i$ (defined hereafter) and modulated by the tSZ frequency spectrum integrated over the frequency response, $\fnu$, in $\mu {\rm K}$ units. The integration along the line-of-sight for $\Tth$ was performed out to $r=5\Rfive$, and $\thetas$ is the scale radius, that is to say the characteristic scale of the cluster. 

For \Planck, $b_i$ is simply the Gaussian beam at frequency $\nu_i$, which is assumed to be azimuthally symmetric. For SPT-SZ, $b_i=B \ast T_i$, where the convolution of the azimuthally symmetric Gaussian beam is represented by $B$ (FWHM=1.75~arcmin) and the filter transfer function is $T_i$ at frequency $\nu_i$. The filter transfer function is not azimuthally symmetric so, in practice, we performed the convolution in all-sky Healpix maps at ten locations centered on the ten SPT clusters with the highest signal-to-noise ratio (S/N), cut tangential maps centered on $b_i$, and averaged them.

Assuming a pressure profile $\Tth$, MMF3 obtains the linear estimate of $\yo$ with minimum variance:
\begin{equation}
\estyo = \int d^2x \; \vec{\Psit}^t(\vec{x}) \cdot \vec{m}(\vec{x}),
\end{equation}
where
\begin{equation}
\vec{\Psit}(\vec{k}) = \sigt^2 \vec{P}^{-1}(\vec{k})\cdot \vec{\tth}(\vec{k}),
\end{equation}
with
\begin{eqnarray}
\label{eq:sigt}
\sigt          & \equiv & \left[\int d^2k\; 
     \vec{\tth}^t(\vec{k})\cdot \vec{P}^{-1} \cdot
     \vec{\tth}(\vec{k})\right]^{-1/2}
\end{eqnarray}
and
\begin{equation}
 \vec{P}(\vec{k}) = <\vec{n}^*(\vec{k}) \vec{n}(\vec{k})> .
\end{equation}
We note that $\vec{k}$ is the two dimensional spatial frequency; and $\vec{P}(\vec{k})$ is the power spectrum matrix of the noise across frequency channels and was estimated directly from the data since the SZ signal is small compared to the other signals. The S/N of the measurement is given by
\begin{equation}
{S \over N}={\estyo \over \sigt}.
\end{equation}

The total integrated flux in the cylinder of radius $r=5\Rfive$ can be estimated as
\begin{equation}
 Y_{5\Rfive} = \estyo \int_{x<5\theta_{500}} d\Omega \Tth(x),
\end{equation}
where $\theta_{500}$ is the angle subtended by $\Rfive$ at the cluster redshift. This total flux was then converted to the flux in the sphere of radius $r=\Rfive$, $Y_{500}$, by multiplying $Y_{5\Rfive}$ by the ratio of ${\int_0^{\Rfive} dr 4 \pi r^2 \tth(r)} \over {\int_{x<5\theta_{500}} d\Omega \Tth(x)}$~\citep[Appendix A of][]{melin2011}. We note that $\tth(r)$ is the three dimensional profile while $\Tth(x)$ is the two dimensional profile, that is, $\tth(r)$ integrated along the line-of-sight.
In the following, we adopt the profile of~\citet{arnaud2010} for $\tth(r)$, except in Section~\ref{sec:spt_in_plck} and Appendix~\ref{sec:sptszphot} in which we adopt a $\beta$-profile to match the SPT-SZ cluster modeling.

We can run MMF3 in unblind or blind mode, that is, by fixing the position and size of the cluster or in letting the position and/or size free. In blind mode, we adopted the size and/or the position that maximize(s) the S/N and refer to them(it) as a blind size (and blind position). The blind flux was estimated by fixing the filter size to the blind size and the position to the blind position if also left free. The blind mode is further described in Sect.~\ref{sec:catalogue_construction}.

\section{Consistency between public products and results published by the SPT and \Planck\ collaborations}
\label{sec:public_consistency}

We checked that the results obtained with SPT filter transfer functions applied to SPT-SZ maps are consistent with SPT published results. We extracted SPT point sources~\citep{mocanu2013} in individual frequency maps using the filter transfer functions and we compared our recovered flux to the published flux. The results are presented in Appendix~\ref{sec:filterfunc}. The filter functions are accurate for point source flux $S<50 \, {\rm mJy}$. For point sources with $S>50 \, {\rm mJy}$, the flux is biased a few percent high: This is expected because the regions surrounding bright point sources are excluded from the fitting of the time stream data~\citep[see Sect. 4.1.1 of][]{chown2018}, so the filter transfer functions are not expected to model these regions of the maps.

We then extracted SPT clusters~\citep{bleem2015} using the SPT filter transfer functions and the SPT-SZ maps. The results are shown in Appendix~\ref{sec:sptszphot}. SPT cluster flux was recovered without any biases for fluxes $Y^{0.75}\, {\rm SPT}<2 \times 10^{-4} \, {\rm arcmin}^2$, while for $Y^{0.75}\, {\rm SPT}>2 \times 10^{-4} \, {\rm arcmin}^2$ there is a few percent bias to larger fluxes\footnote{$Y^{0.75}\, {\rm SPT}$ is the integrated SPT flux in a cylinder of radius 0.75~arcmin}. This bias may come from the filter transfer function, which is biased high for bright sources, as shown above. The SPT cluster size was recovered without any biases, and SPT S/N was recovered 10\% low with respect to the results published by the SPT collaboration. 
The SPT public maps~\citep{chown2018} are shallower than the maps used in~\cite{bleem2015}. We suspect that the S/N bias to lower values is due to the difference in map depths.
We would expect this bias to disappear if the MMFs were applied to the same SPT data as used in~\cite{bleem2015}.

Finally, we checked the consistency of the \Planck\ cluster properties extracted from the \Planck\ public data to the flux published by the \Planck\ collaboration. The results are presented in Appendix~\ref{sec:plckphot}. We used one of the algorithms developed in the \Planck\ collaboration but we do not expect to find a one-to-one relation between our recovered flux, size and S/N, and the quantities published by the \Planck\ collaboration. This is because we upgraded the maps from $N_{side}=2048$ to 8192, and we changed the coordinate system from Galactic to equatorial. In particular, the change of coordinates modifies the estimation of the noise power spectrum $\vec{P}(\vec{k})$, introducing some scatter in the recovered quantities with respect to the published values, but no significant bias.

We conclude that our extraction method provides results that are consistent with the results published by both the SPT and \Planck\ collaboration.  After these consistency tests between public products and the results published by the collaborations, we checked for consistency across the two data sets.

\section{Consistency between the SPT-SZ and \Planck\ data sets}
\label{sec:sptplanck_consistency}

We checked the consistency between the two data sets by extracting the SPT-SZ clusters in \Planck\ data adopting the SPT cluster modeling and vice versa, that is, by extracting the \Planck\ clusters in SPT-SZ data adopting the \Planck\ cluster modeling.

\subsection{SPT-SZ cluster flux in \Planck\ data}
\label{sec:spt_in_plck}

\begin{figure}
\centering
\includegraphics[width=\hsize]{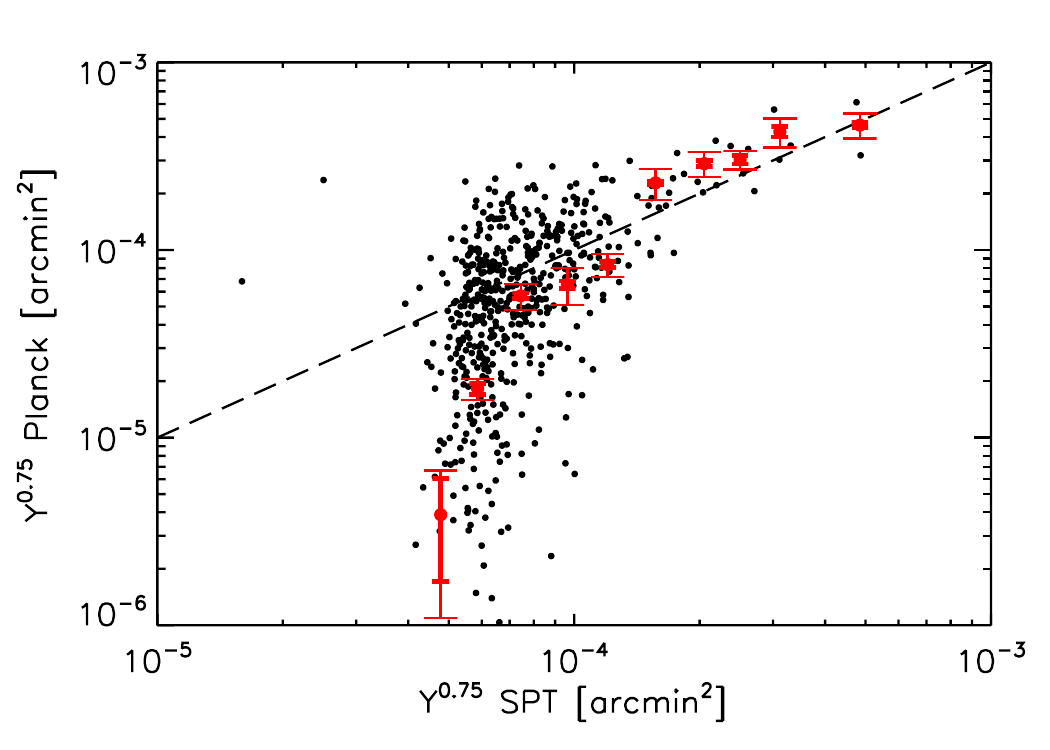}
\caption{\Planck\ flux of the SPT-SZ clusters~\citep{bleem2015} as a function of the published SPT-SZ flux. The black dots are individual clusters. The red diamonds are weighted averages. The thick error bars show statistical errors and the thin error bars were obtained by the bootstrap method. Despite a large scatter, the agreement between the SPT-SZ flux and the \Planck\ flux is good. The drop-off at low flux is the Malmquist bias of the SPT-SZ flux due to the SPT detection threshold.}
              \label{fig:plck_vs_spt_spt}
\end{figure}

We used the full ($\xi >4.5$) SPT-SZ published cluster catalog~\citep{bleem2015}. For the work described in this section, we adopted the $\beta$-profile for the cluster template with the same parametrization as in the SPT-SZ analysis. We applied MMF3 at the location of the clusters, fixing the size $\theta_c$ to the value published in the catalog. The estimated central Compton parameter $\estyo$ was then converted to the integrated flux in a cylinder of radius 0.75~arcmin, $Y^{0.75}$. The flux can be directly compared to the flux given in the published catalog. The results are shown in Fig.~\ref{fig:plck_vs_spt_spt}. 

There is a large dispersion between the individual measured fluxes and the published flux (black dots). This is expected because the \Planck\ maps are noisier than the SPT-SZ maps. We averaged the \Planck\ fluxes in SPT-SZ flux bins (red diamonds). The averaged bin flux is generally in good agreement with the SPT-SZ flux. The agreement is very good at large values ($Y^{0.75} \, {\rm SPT} > 2 \times 10^{-4} \, {\rm arcmin}^2$), and the average \Planck\ flux starts to deviate to lower values with decreasing SPT-SZ flux ($Y^{0.75} \, {\rm SPT} < 2 \times 10^{-4} \, {\rm arcmin}^2$). We attribute this deviation to the Malmquist bias of the SPT-SZ flux due to the SPT detection threshold. Although \Planck\ is less sensitive than SPT, it is not subject to the selection bias of the SPT sample. We conclude that the SPT cluster flux is consistent with the \Planck\ data. 

\subsection{\Planck\ cluster flux in SPT-SZ data}

\begin{figure}
\centering
\includegraphics[width=\hsize]{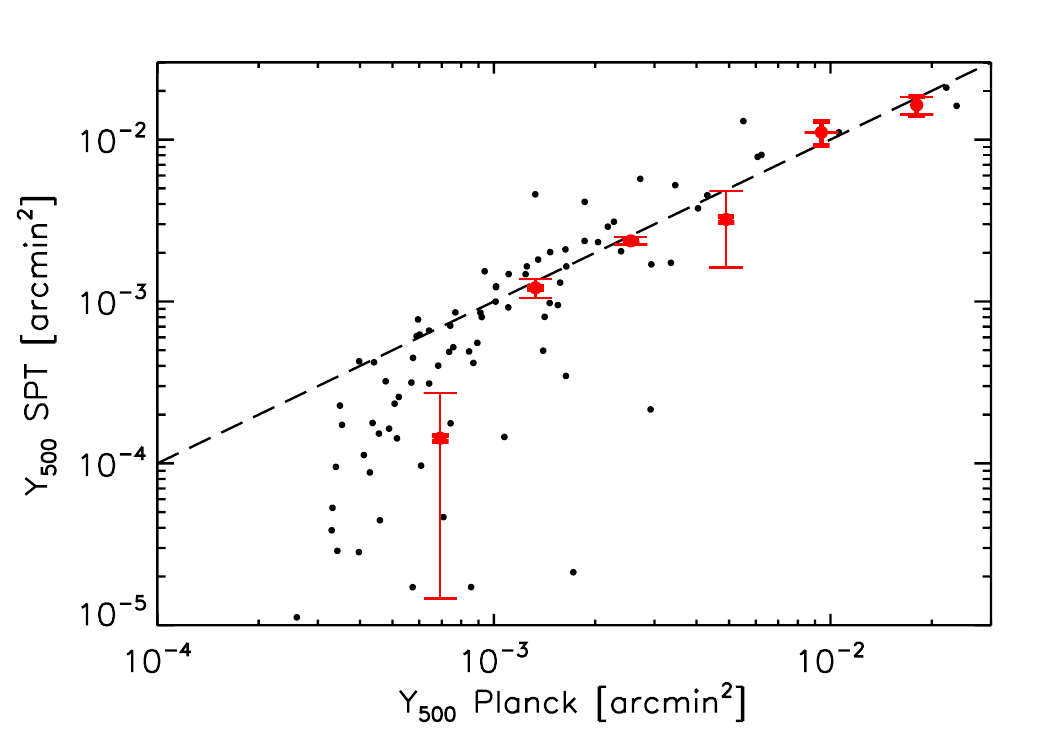}
\caption{SPT fluxes of \Planck\ clusters~\citep[MMF3 sample of PSZ2,][]{PSZ2} as a function of published \Planck\ flux. The agreement between the \Planck\ flux and the SPT-SZ flux is good, as for the SPT sample shown in Fig.~\ref{fig:plck_vs_spt_spt}. The drop-off at low flux is the Malmquist bias of the \Planck\ flux due to the \Planck\ detection threshold. The legends are similar to Fig.~\ref{fig:plck_vs_spt_spt}.}
              \label{fig:spt_vs_plck_plck}
\end{figure}

We used the published \Planck\ catalog PSZ2~\citep[][]{PSZ2}. We restricted the sample to clusters detected by MMF3 ($S/N>4.5$) and used the blind \Planck\ flux and size, that is, the flux and size given at the maximum of the degeneracy contours provided by the \Planck\ collaboration. We extracted the cluster flux from SPT-SZ maps fixing the position to the \Planck\ blind position and the size to the \Planck\ blind size.  The results are given in Fig.~\ref{fig:spt_vs_plck_plck}. The figure shows a large scatter, but no significant bias for $Y_{500} \, Planck > 10^{-3} \, {\rm arcmin}^2$.  At lower \Planck\ fluxes ($Y_{500} \, Planck < 10^{-3} \, {\rm arcmin}^2$), the SPT-SZ flux deviates toward smaller values. This deviation is likely due to the Malmquist bias in the \Planck\ fluxes at the \Planck\ detection threshold. We conclude that the \Planck\ cluster flux is consistent with the SPT-SZ data.

With the two data sets being consistent, we then applied the MMF3 algorithm jointly to the \Planck\ and SPT-SZ maps.

\section{The PSZSPT cluster catalog}
\label{sec:pszspt}

We first describe the construction of our candidate list using MMF3 in Sect.~\ref{sec:catalogue_construction}. We match our candidates to known clusters in Sect.~\ref{sec:identification}. We check for missed SPT and PSZ2 clusters in Sect.~\ref{sec:missed}.  Finally, we estimate the completeness of our catalog in Sect.~\ref{sec:completeness}. The format and fields of the PSZSPT catalog are given in Appendix~\ref{sec:description}.

\subsection{Construction of the catalog}
\label{sec:catalogue_construction}

We divided the \Planck\ and SPT-SZ Healpix maps into 52 overlapping tangential maps of 10x10 deg$^2$ (pixel size = 0.43~arcmin), covering the SPT-SZ footprint, and we ran the MMF3 algorithm blindly.
The description of the blind MMF3 algorithm is given in~\cite{melin2012}, \cite{ESZ}, \cite{PSZ1} and \cite{PSZ2}. We briefly recall here its main features and we give the differences with the implementation used for the \Planck\ analyses.\\

The algorithm was run blindly on the individual maps. We fixed a detection threshold $q_{\rm thres}$. We filtered each map with a set of 32 logarithmically spaced sizes $\thetas$, ranging from 0.8 to 30 arcmin. We looked for the maximum in the filtered maps corresponding to our first cluster candidate. We masked it and looked for the second maxima. We continued until there were no more remaining maxima above the detection threshold. In doing so, we built a blind catalog for each tangential map, which includes blind positions (corresponding to the position of the maxima), blind sizes (corresponding to the sizes maximizing the S/N), and blind fluxes (given by the filter output at the blind positions and for the blind sizes). We proceeded similarly for the 52 tangential maps. We then constructed a catalog from the 52 individual catalogs by merging detections with separation less than 2.5 arcmin. We then proceeded with a second pass of the algorithm. We divided the \Planck\ and SPT-SZ Healpix maps in tangential maps centered on the first pass detections and ran the algorithm again. The second pass allowed us to obtain better estimates for the position, size, and flux, and to reject detections with refined S/N lower than $q_{\rm thres}$.

The \Planck\ and SPT-SZ maps include bright point sources which must be masked to avoid spurious detections. We implemented the same methodology for the SPT-SZ and \Planck\ tangential maps. We used a single frequency matched filter for each map and we detected point sources with $S/N>8$. The point source detection was performed in each of the three SPT-SZ maps and each of the six \Planck\ maps. For the SPT-SZ data, we masked circular regions of 5~arcmin radius around each point source and rejected any SZ detection in a 7.5~arcmin radius. For comparison, \cite{bleem2020} mask in 4~arcmin radius for $S/N>5$ point sources and rejected detections within 8~arcmin radius. For the \Planck\ data, we masked in 10~arcmin radius and rejected detections in 15~arcmin radius because of the larger beams.

In summary, the differences between this implementation of MMF3 and the implementation used for the official \Planck\ catalogs are as follows: the partial sky coverage (SPT-SZ footprint instead of all-sky); the pixel size of the maps (0.43~arcmin instead of 1.72~arcmin); the orientation of the tangential maps (equatorial pole to the north instead of the Galactic pole to the north); the filter transfer function for the SPT-SZ maps; the merging separation (2.5~arcmin instead of 10~arcmin); and the removal of bright point sources (made for SPT-SZ and \Planck\ data on tangential maps instead of all-sky maps for \Planck).

We fixed the detection threshold $q_{\rm thres}$ to 5 for our joint catalog. We additionally applied the SPT-SZ boundary mask and the \Planck\ cluster union mask (which keeps the 85\% cleanest part of the sky).
We thus obtain a catalog of 419 detections. In Sect.~\ref{sec:identification}, we identify our detections with known clusters, and we present the completeness of our catalog in Sect.~\ref{sec:completeness}.

\subsection{Identification of known clusters}
\label{sec:identification}

We followed the methodology of~\cite{tarrio2019} to identify clusters in our joint catalog. We first identified clusters in the SPT-SZ catalog~\citep{bleem2015} (Sect.~\ref{sec:x_spt}) and in the PSZ2 catalog~\citep{PSZ2} (Sect.~\ref{sec:x_plck}). We then cross matched the catalog with other relevant catalogs in the SPT-SZ footprint (Sect.~\ref{sec:x_others}): the Meta-Catalogue of X-ray Clusters MCXC~\citep{piffaretti2011}, the Meta-Catalogue of SZ Clusters MCSZ\footnote{\url{https://www.galaxyclusterdb.eu/}}, the ComPRASS catalog~\citep{tarrio2019}, the Abell catalog~\citep{abell1989}, the cluster catalog from the Blanco Cosmology Survey~\citep{desai2012, bleem_blanco2015}, the MARD-Y3 catalog~\citep{klein2019}, and the WHY catalog~\citep{wen2018}. We finally used SIMBAD and NED (Sect.~\ref{sec:x_nedsimbad}). We present the unidentified detections in Sect.~\ref{sec:x_unmatch}.

\subsubsection{SPT-SZ catalog}
\label{sec:x_spt}

First, we matched each of the 419 blind detections to the closest cluster of the SPT-SZ cluster catalog. We then separated the detections into two sets: the detections matched to a SPT-SZ cluster with an estimated mass $M_{500}$ and redshift $z$, and the detections matched to a SPT-SZ detection without an estimated mass.

\begin{figure}
\centering
\includegraphics[width=\hsize]{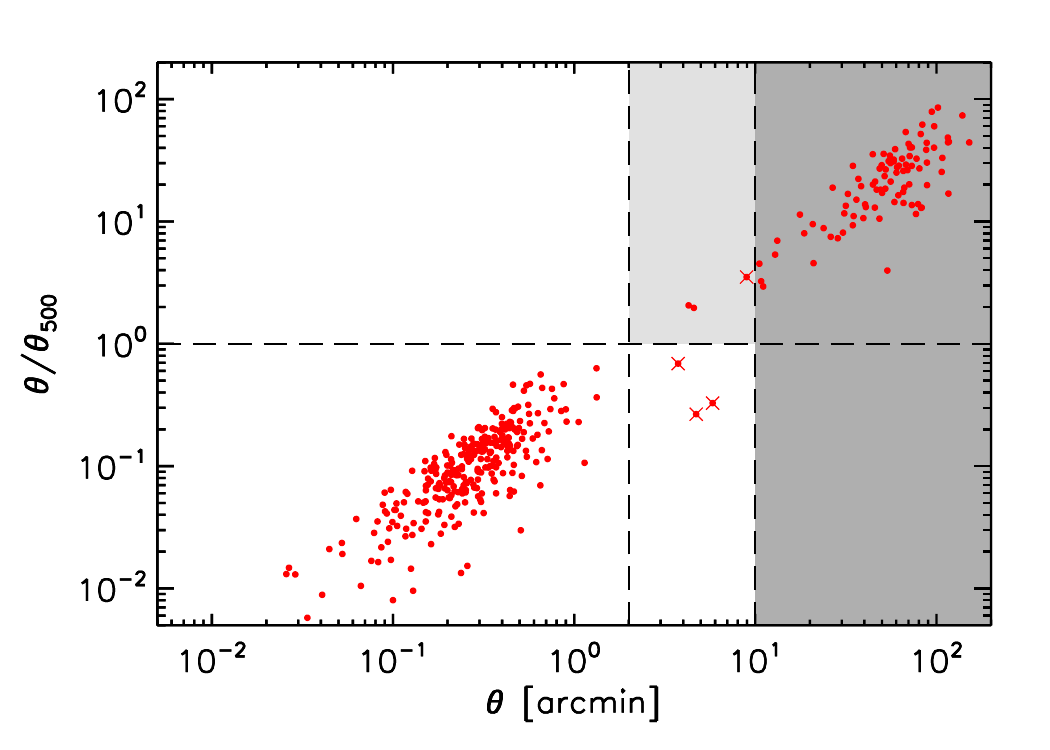}
\caption{Blind detections matching SPT-SZ clusters with published mass $M_{500}$ and redshift $z$. Candidates in the white (dark gray) area are matched (not matched) to SPT-SZ clusters. Candidates in the light gray area are possibly matched. The crosses mark multiple detections which are matched to SPT-SZ clusters already associated with a closer detection, and they are thus not considered as being matched to an SPT-SZ cluster.}
\label{fig:sptsz_match_withz}
\end{figure}

We plotted the first set in the $\theta/\theta_{500}$ versus $\theta$ plane where $\theta$ is the distance between the blind candidate and the closest SPT-SZ cluster and $\theta_{500}$ is the angular radius corresponding to the published SPT-SZ mass $M_{500}$ and redshift $z$. The result is given in Fig.~\ref{fig:sptsz_match_withz}, which shows two clouds of detections: the detections which can be matched to an SPT-SZ cluster in the lower left corner and the detections which cannot be matched in the upper right corner. We define a detection as being matched to an SPT-SZ cluster if $\theta<2\, {\rm arcmin}$, or $\theta/\theta_{500}<1$ and $\theta<10\, {\rm arcmin}$. We label it as rank=1. These detections correspond to the white region. We also define a detection as not being matched to a SPT-SZ cluster if $\theta>10\, {\rm arcmin,}$ corresponding to the dark gray regions. We note it as rank=0. We then define the light gray region ($\theta/\theta_{500}>1$ and $2\, {\rm arcmin} < \theta<10\, {\rm arcmin}$) as the possibly matched clusters (rank=2).

\begin{figure}
\centering
\includegraphics[width=\hsize]{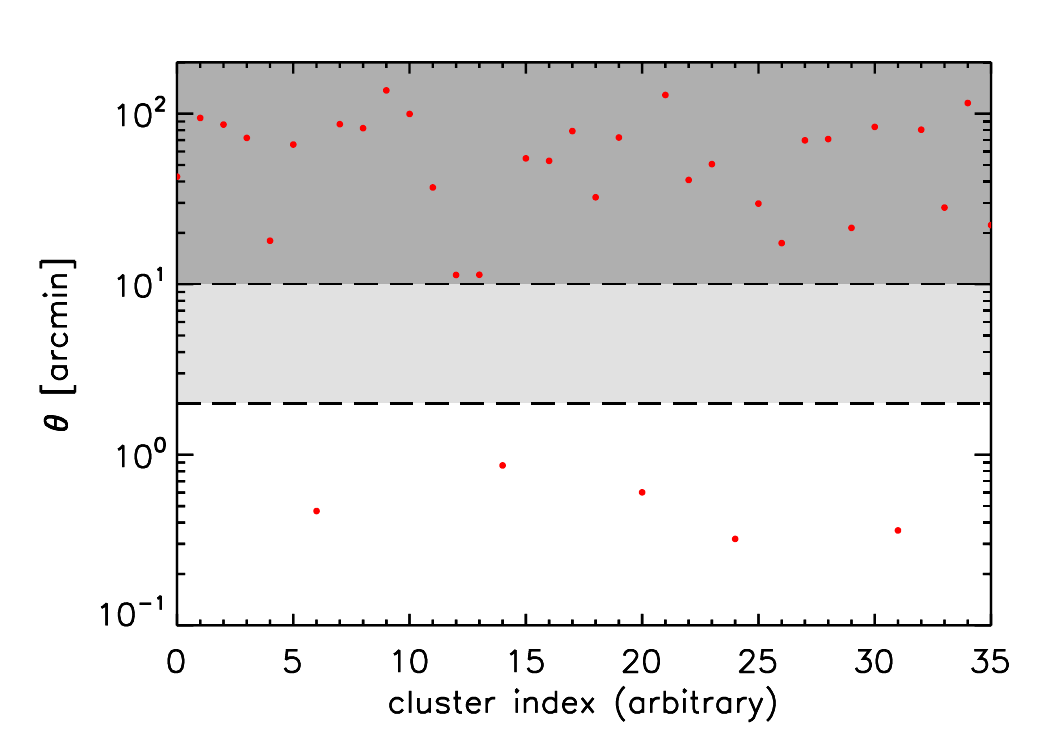}
\caption{Blind detections matching SPT-SZ clusters without published mass $M_{500}$. As for Fig.~\ref{fig:sptsz_match_withz}, candidates in the white (dark gray) area are matched (not matched) to SPT-SZ clusters.}
\label{fig:sptsz_match_noz}
\end{figure}

For the second set without an estimated mass, we associated clusters (rank=1) if $\theta<2\, {\rm arcmin}$ and we did not associate clusters (rank=0) if $\theta>10\, {\rm arcmin}$. The detections with $2\, {\rm arcmin} < \theta<10\, {\rm arcmin}$ were set as possibly associated (rank=2). The association is shown in Fig.~\ref{fig:sptsz_match_noz}. We note that there is no rank=2 detection of this category (light gray area) for the matching with the SPT catalog.

Finally, we checked for detections that were matched or possibly matched (rank=1 or rank=2)  to the same SPT-SZ cluster, and we kept only the closest associated detection, giving priority to rank=1 over rank=2. We degraded the other multiple associations from rank=1 or rank=2 to rank=3. They are marked as crosses in Fig.~\ref{fig:sptsz_match_withz}. There are no rank=3 detections in Fig.~\ref{fig:sptsz_match_noz}. We investigate rank=3 detections further using SIMBAD and NED in Sect.~\ref{sec:x_nedsimbad}.

\begin{figure}
\centering
\includegraphics[width=\hsize]{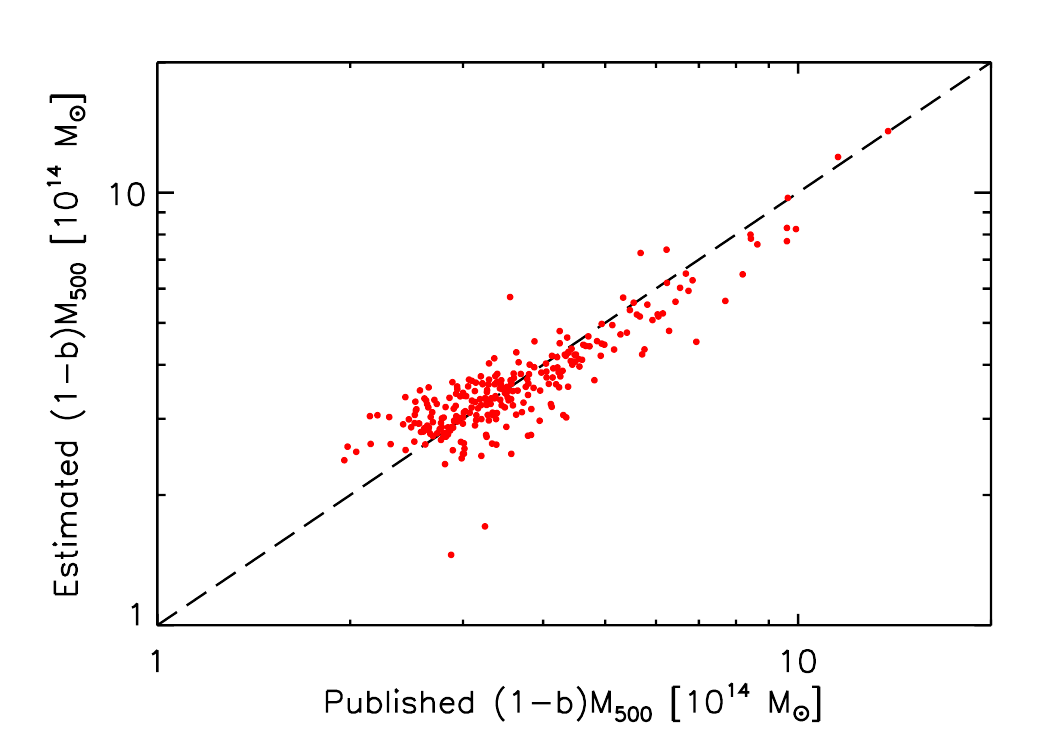}
\caption{Extracted masses versus published SPT-SZ masses for the joint detections matched to SPT-SZ clusters. Each point is a detection matched to an SPT-SZ cluster. We note that $(1-b)$ is the "mass bias factor" which relates the true mass $M_{500}$ to the {\em XMM-Newton}-like mass $M_{500,X}=(1-b)M_{500}$.}
\label{fig:sptsz_match_mass}
\end{figure}

As a final test, we estimated the mass for rank=1 associations using the cluster redshift $z$ and a X-ray prior on the $Y-M$ scaling relation \citep[see][]{PSZ1,PSZ2,tarrio2019}. We plotted it against the published SPT-SZ mass after recalibrating it by a factor of 0.8 to account for our mass definition, as was done in~\cite{tarrio2019} and as implemented in the MCSZ catalog. The result is shown in Fig.~\ref{fig:sptsz_match_mass}. The agreement between the two masses is good. We discuss this figure further, including outliers, in Appendix~\ref{sec:masscomp}.

Among the 419 joint detections, we found 290 matched to SPT-SZ clusters (rank=1) and two possibly matched to SPT-SZ clusters (rank=2).

\subsubsection{PSZ2 catalog}
\label{sec:x_plck}

\begin{figure}
\centering
\includegraphics[width=\hsize]{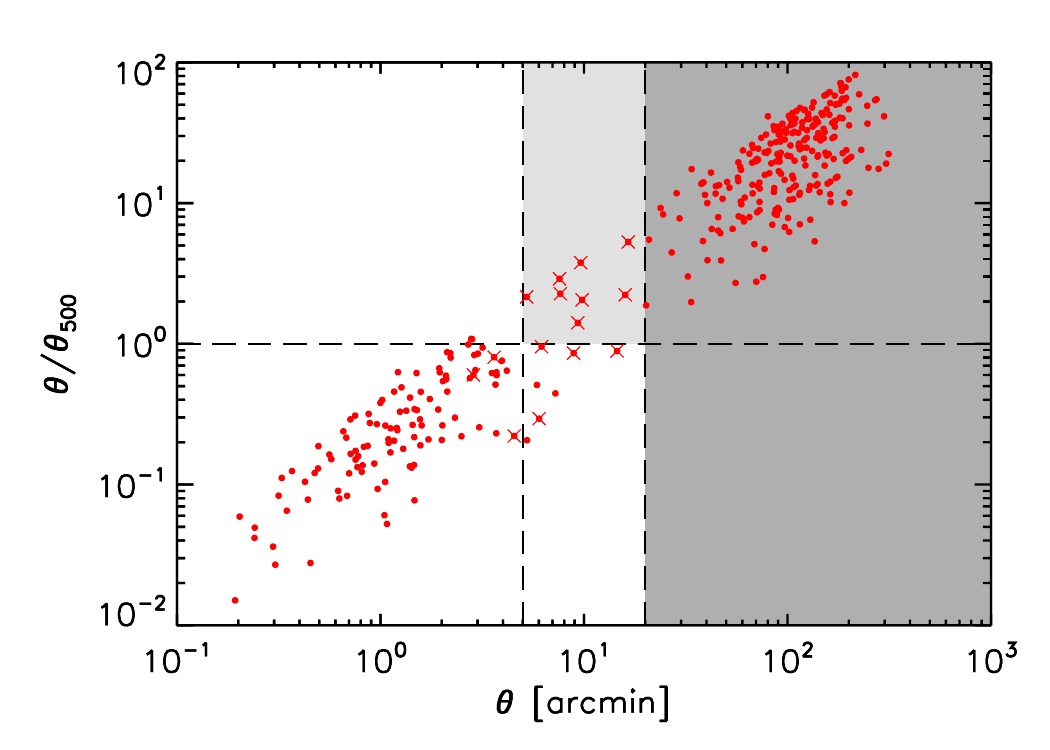}
\caption{Blind detections matching PSZ2 clusters with published mass $M_{500}$ and redshift $z$. The legend is the same as Fig.~\ref{fig:sptsz_match_withz}.}
\label{fig:psz2_match_withz}
\end{figure}

\begin{figure}
\centering
\includegraphics[width=\hsize]{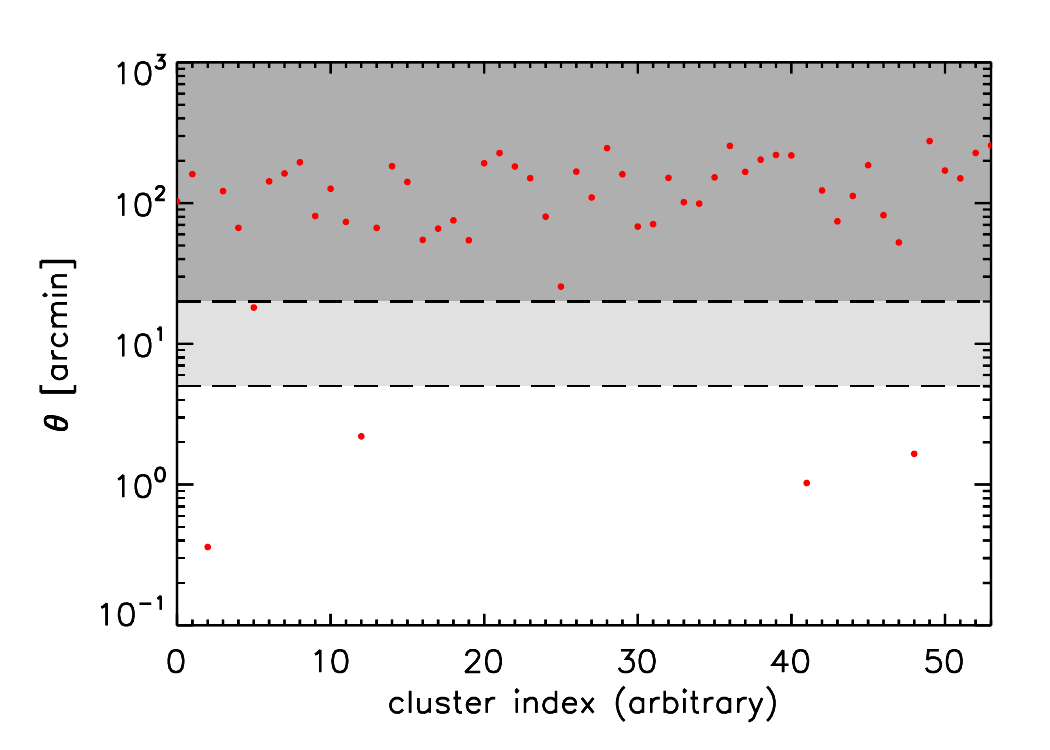}
\caption{Blind detections matching PSZ2 clusters without published mass $M_{500}$. As for Fig.~\ref{fig:sptsz_match_noz}, candidates in the white (dark gray) area are matched (not matched) to PSZ2 clusters. Candidates in the light gray region are possibly matched to a PSZ2 cluster.}
\label{fig:psz2_match_noz}
\end{figure}

We applied the same methodology to the PSZ2 cluster catalog. The matching with PSZ2 clusters with mass $M_{500}$ and redshift $z$ is shown in Fig.~\ref{fig:psz2_match_withz}, and the matching with clusters without masses is shown in Fig.~\ref{fig:psz2_match_noz}. Given the larger beam size of \Planck\, we changed the limits for  association. We define a detection as being matched to a PSZ2 cluster if $\theta<5\, {\rm arcmin}$, or $\theta/\theta_{500}<1$ and $\theta<20\, {\rm arcmin}$ (white region in Fig. ~\ref{fig:psz2_match_withz}, rank=1). A detection with $\theta>20\, {\rm arcmin}$ (dark gray region, rank=0) is considered as not being matched, and the detection with $\theta/\theta_{500}>1$ and $5\, {\rm arcmin} < \theta<20\, {\rm arcmin}$ (light gray region, rank=2) is considered as being possibly matched.

The association with PSZ2 clusters without a published mass is shown in Fig.~\ref{fig:psz2_match_noz} with the same color coding of the regions for matched, not matched, and possibly matched detections.
We also discarded double matching as we did for SPT-SZ (rank=3). 

\begin{figure}
\centering
\includegraphics[width=\hsize]{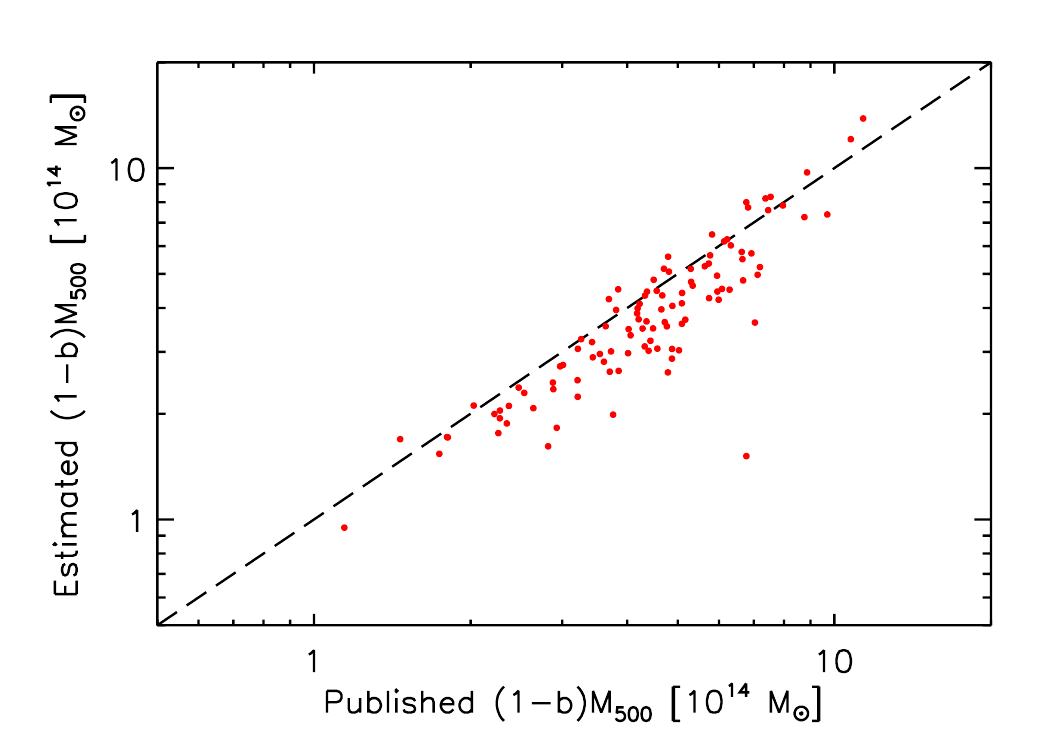}
\caption{Extracted masses versus published PSZ2 masses for the joint detections matched to PSZ2 clusters. Each point is a detection matched to a PSZ2 cluster. We note that $(1-b)$ is the same mass bias factor as in Fig.~\ref{fig:sptsz_match_mass}.}
\label{fig:psz2_match_mass}
\end{figure}

As for the SPT-SZ matching, we estimated the mass for rank=1 associations and plotted them against the published PSZ2 masses in Fig.~\ref{fig:psz2_match_mass}. The agreement is also good. We discuss this figure further, including outliers and the systematic underestimation of the PSZ2 mass, in Appendix~\ref{sec:masscomp}.

Among the 419 joint detections, we found 112 matched to PSZ2 clusters (rank=1) and one possibly matched to PSZ2 clusters (rank=2).
Finally, 82 detections are matched or possibly matched to a SPT-SZ and a PSZ2 cluster at the same time.
At this stage, we thus have 419-292-113+82=96 detections, which are not matched or possibly matched to a cluster from the SPT-SZ or the PSZ2 catalog.

\subsubsection{Other catalogs}
\label{sec:x_others}

After the matching with the SPT-SZ and the PSZ2 catalogs, we proceeded with the same methodology for the MCSZ Meta catalog, the MCXC Meta catalog~\citep{piffaretti2011}, the ComPRASS catalog~\citep{tarrio2019}, the Abell Southern catalog~\citep[Table 4 and 5 of][]{abell1989}, the Blanco Cosmology Survey~\citep[BCS,][]{desai2012, bleem_blanco2015}, the MARD-Y3 catalog~\citep{klein2019}, and the WHY catalog~\citep{wen2018}. For the matching procedure, we adopted the same values as for the PSZ2 for all of these catalogs (5 and 20~arcmin), except for the SPT clusters in the MCSZ for consistency with Sect.~\ref{sec:x_spt}, for which we adopted the SPT-SZ values (2 and 10~arcmin). For the MARD-Y3 catalog, we excluded multiple detections of the same source (by fixing fmult=0), and we set the contamination cut to 5\% (by fixing fcontlfcr<0.05). For the WHY catalog, we used clusters with richness $R_{L*} \geqslant 30$ and the mass-richness relation from~\cite{wen2015}.

Among the 96 detections, which are not matched or possibly matched to a SPT-SZ or a PSZ2 cluster, we find that:
\begin{itemize}
\item three are matched or possibly matched to MCSZ clusters
\begin{itemize}
\item PLCK G260.7-26.3 at z=0.68,
\item PSZ1 G295.98-69.26, and
\item PSZ1 G352.42-48.31;
\end{itemize}
\item three are matched or possibly matched to MCXC clusters
\begin{itemize}
\item MCXC J0330.0-5235 at z=0.0624,
\item MCXC J0245.2-4627 at z=0.0868, and
\item MCXC J2321.5-4153 at z=0.0894;
\end{itemize}
\item ten are matched or possibly matched to ComPRASS clusters
\begin{itemize}
\item PSZRX G264.82-51.13 at z=0.0624 is also MCXC J0330.0-5235,
\item PSZRX G348.32-66.45 at z=0.0894 is also MCXC J2321.5-4153, 
\item PSZRX G345.22-32.90 at z=0.237,
\item PSZRX G271.53-56.58 at z=0.3,
\item PSZRX G252.12-34.15,
\item PSZRX G260.92-35.30,
\item PSZRX G282.53-53.69,
\item PSZRX G282.66-54.84,
\item PSZRX G249.39-34.22, and
\item PSZRX G352.49-33.18;
\end{itemize}
\item 16 are matched or possibly matched to Abell clusters
\begin{itemize}
\item Abell S854 at z=0.0561,
\item Abell 3128 at z=0.0624 is also PSZRX G264.82-51.13 and MCXC J0330.0-5235,
\item Abell 3047 at z=0.0868 is also MCXC J0245.2-4627,
\item Abell 3998 at z=0.0894 is also PSZRX G348.32-66.45 and MCXC J2321.5-4153,
\item Abell 3279 at z=0.1425,
\item Abell 3665 at z=0.237 is also PSZRX G345.22-32.90,
\item Abell S295 at z=0.3 is also PSZRX G271.53-56.58,
\item Abell S526 is also PSZRX G252.12-34.15,
\item Abell 3886,
\item Abell S47,
\item Abell S132,
\item Abell S184,
\item Abell 3209,
\item Abell 3236,
\item Abell 3818, and
\item Abell S1089;
\end{itemize}
\item two are matched or possibly matched to BCS clusters
\begin{itemize}
\item BCS J233151-5736.2 at z=0.27 and
\item BCS J051723-5325.5;
\end{itemize}
\item 23 are matched or possibly matched to MARD-Y3 clusters
\begin{itemize}
\item MARD J201247.7-565058 at z=0.0514,
\item MARD J032959.4-523546 at z=0.0624 is also Abell 3128, PSZRX G264.82-51.13, and MCXC J0330.0-5235,
\item MARD J052449.4-613521 at z=0.0788,
\item MARD J055700.0-424702 at z=0.0825,
\item MARD J024524.4-462750 at z=0.0868 is also Abell 3047 and MCXC J0245.2-4627,
\item MARD J232122.2-415047 at z=0.0894 is also Abell 3998, PSZRX G348.32-66.45, and MCXC J2321.5-4153,
\item MARD J211813.3-474018 at z=0.1103,
\item MARD J203848.2-463207 at z=0.1300,
\item MARD J043805.0-455901 at z=0.1425 is also Abell 3279,
\item MARD J200938.7-531030 at z=0.237 is also Abell 3665 and PSZRX G345.22-32.90,
\item MARD J021020.5-464352 at z=0.292,
\item MARD J024528.5-530203 at z=0.3 is also Abell S295 and PSZRX G271.53-56.58,
\item MARD J044052.4-474315 at z=0.3075,
\item MARD J011511.6-595828 at z=0.3972,
\item MARD J223243.0-542921 at z=0.4248,
\item MARD J215445.0-593625 at z=0.4317,
\item MARD J043431.7-572717 at z=0.4709,
\item MARD J061633.0-522732 at z=0.68 is also PLCK G260.7-26.3,
\item MARD J052243.5-461106,
\item MARD J014307.6-582601,
\item MARD J022718.3-592935,
\item MARD J022133.6-583730, and
\item MARD J040244.4-533049; and
\end{itemize}
\item 31 are matched or possibly matched to WHY clusters
\begin{itemize}
\item J201358.6-570223 at z=0.0514 is also MARD J201247.7-565058,
\item J201050.6-564025 at z=0.0561 is also Abell S854,
\item J032950.6-523447 at z=0.0624 is also MARD J032959.4-523546, Abell 3128, PSZRX G264.82-51.13, and MCXC J0330.0-5235,
\item WHY J052352.8-614054 at z=0.0788 is also MARDJ052449.4-613521,
\item J024513.3-462719 at z=0.0868 is also MARD J024524.4-462750, Abell 3047, and MCXC J0245.2-4627,
\item J042953.7-463447 at z=0.1389,
\item J043815.2-455712 at z=0.1425 is also MARD J043805.0-455901 and Abell 3279,
\item J223249.9-545158 at z=0.151,
\item WHY J201309.0-463059 at z=0.1746,
\item WHY J215446.7-461610 at z=0.2161,
\item WHY J214150.7-481507 at z=0.2319,
\item J200950.6-530948 at z=0.237 is also MARD J200938.7-531030, Abell 3665, and PSZRX G345.22-32.90,
\item WHY J060809.4-615348 at z=0.2424,
\item WHY J010643.2-575952 at z=0.2568,
\item WHY J064816.1-611730 at z=0.2707,
\item WHY J220455.6-622321 at z=0.2761,
\item WHY J061026.5-461946 at z=0.2782,
\item J000314.6-525516 at z=0.2799,
\item WHY J021011.3-464254 at z=0.292 is also MARD J021020.5-464352,
\item WHY J044100.6-474229 at z=0.3075 is also MARD J044052.4-474315,
\item WHY J234242.1-465853 at z=0.3322,
\item WHY J212830.6-463750 at z=0.3381,
\item J041355.6-461148,
\item J002651.3-514159,
\item J052316.1-460241,
\item J230911.4-474414,
\item J040244.8-533313,
\item WHY J052112.2-435700,
\item WHY J014308.1-582621,
\item WHY J022131.4-583722, and
\item WHY J223303.5-542946.
\end{itemize}
\end{itemize}
These matched detections are mainly at low ($z<0.1$) and intermediate ($z\sim0.3$) redshift, or have no published redshift. Some of them are in common between the catalogs. After this new set of associations, 48 detections among the 96 remain unmatched. Among these 48 detections, two are rank=3 detections (for which we broke a rank=1 and rank=2 association) and the 46 others are not classified.

\subsubsection{SIMBAD and NED}
\label{sec:x_nedsimbad}

We searched for counterparts in SIMBAD and NED for the remaining 48 detections, which are not matched to any of the clusters in the studied catalogs. We set the search radius to 20~arcmin for the two databases and we looked for galaxy cluster-type objects.
We did not find a cluster in the search radius for 34 detections. We set rank=0 (meaning unidentified) for these 34 detections. 

For the other 48-34=14 detections, we found two obvious bright and large clusters (ACO~S~1063 and ACO~3911) close to the two rank=3 detections, confirming that these two detections are multiple (therefore false) detections produced by the algorithm. We also found found SPT clusters that are not included in the cluster catalog provided by~\cite{bleem2015}: Three were excluded from the official SPT catalog because they are close to a bright point source~\citep[Table 3 of~][]{bleem2015}, one is in~\cite{saro2015}, but not in \cite{bleem2015}. We set rank=1 for these four detections. The remaining 14-2-4=8 detections are not obviously matched to the clusters found in the search radius because the counterparts are at a distance greater than 7~arcmin. We thus set rank=0 (unidentified) for these eight detections. We additionally checked for detections in the PSZSPT catalog separated by less than 5 arcmin and noticed that PSZSPT J2012-5649 and PSZSPT J2012-5652 are associated with clusters located at the same redshift. After a SIMBAD and NED search at the location of the two clusters, we noticed that PSZSPT J2012-5652 might actually be a substructure of PSZSPT J2012-5649 (ACO~3667). We thus set PSZSPT J2012-5652 as a rank=3 (multiple) detection.

In summary, after the SIMBAD and NED search, there were seven additional identifications (three multiple rank=3 clusters, four SPT rank=1 clusters not in~\cite{bleem2015}) and 34+8=42 unidentified (rank=0 detections).

\subsubsection{Unidentified detections}
\label{sec:x_unmatch}

\begin{figure}
\centering
\includegraphics[width=0.75\hsize]{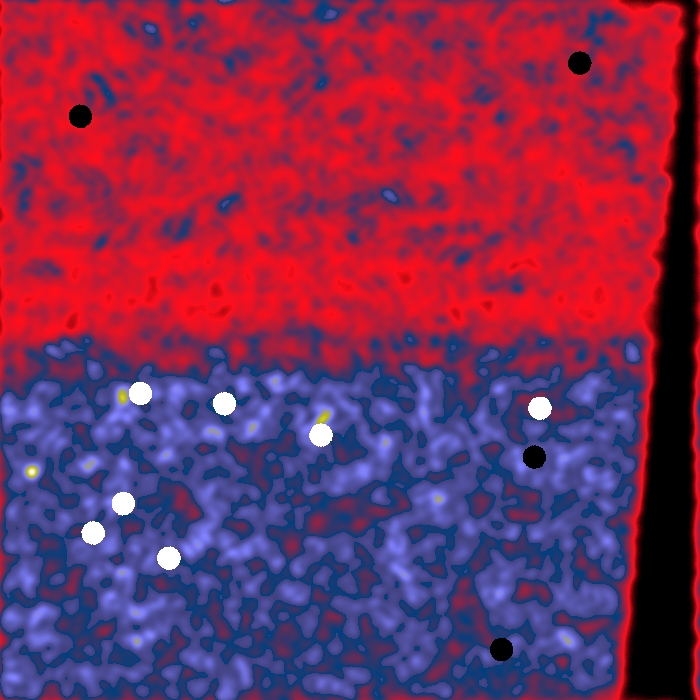}
\caption{Identified (black disks) and unidentified (white disks) detections in the $10 \times 10 \deg^2$ tangential map centered on ra=307.364~deg and dec=-45.6752~deg. The background image displays the local variance of the instrumental noise at the SPT-SZ 150~GHz frequency, which is higher in the bottom part of the map.}
\label{fig:unidentified}
\end{figure}

We looked for the spatial distribution of the 42 unidentified detections in the SPT footprint. We found no specific pattern which could indicate a problem with the extraction algorithm related to the possible systematics in the maps, except in a specific location on the edge of the SPT footprint. This location is displayed in Fig.~\ref{fig:unidentified}. The figure shows the local variance of the instrumental SPT-SZ noise at 150~GHz as the half map difference $\Delta HM$ squared filtered by a 10~arcmin FWHM Gaussian beam $G$ minus the square of the filtered half map difference:
\begin{equation}
\int d\vec{x'} G(\vec{x-x'}) \Delta HM^2(\vec{x'}) - \left ( \int d\vec{x'} G(\vec{x-x'}) \Delta HM(\vec{x'}) \right )^2
.\end{equation}
The black triangle on the right of the image is outside the SPT footprint. There is a clear separation between the top and bottom of the image with the upper part of the image being less noisy than the bottom part (factor 2.5 between the variances of the two parts). This noise difference is due to different integration times or instrumental sensitivities around this location. The detections are displayed as disks. The black and white circles are identified and unidentified detections, respectively. A cluster of unidentified detections is visible in the bottom part. The MMF algorithm estimates the noise on the full map. It is thus possible that the noise is not correctly estimated in this specific patch due to the inhomogeneity of the SPT-SZ instrumental noise. However, four detections have been identified with known clusters in this patch, two of them being in the top part and the other two in the bottom part. We thus do not flag unidentified detections in this patch.\\

\begin{figure}
\centering
\includegraphics[width=0.9\hsize]{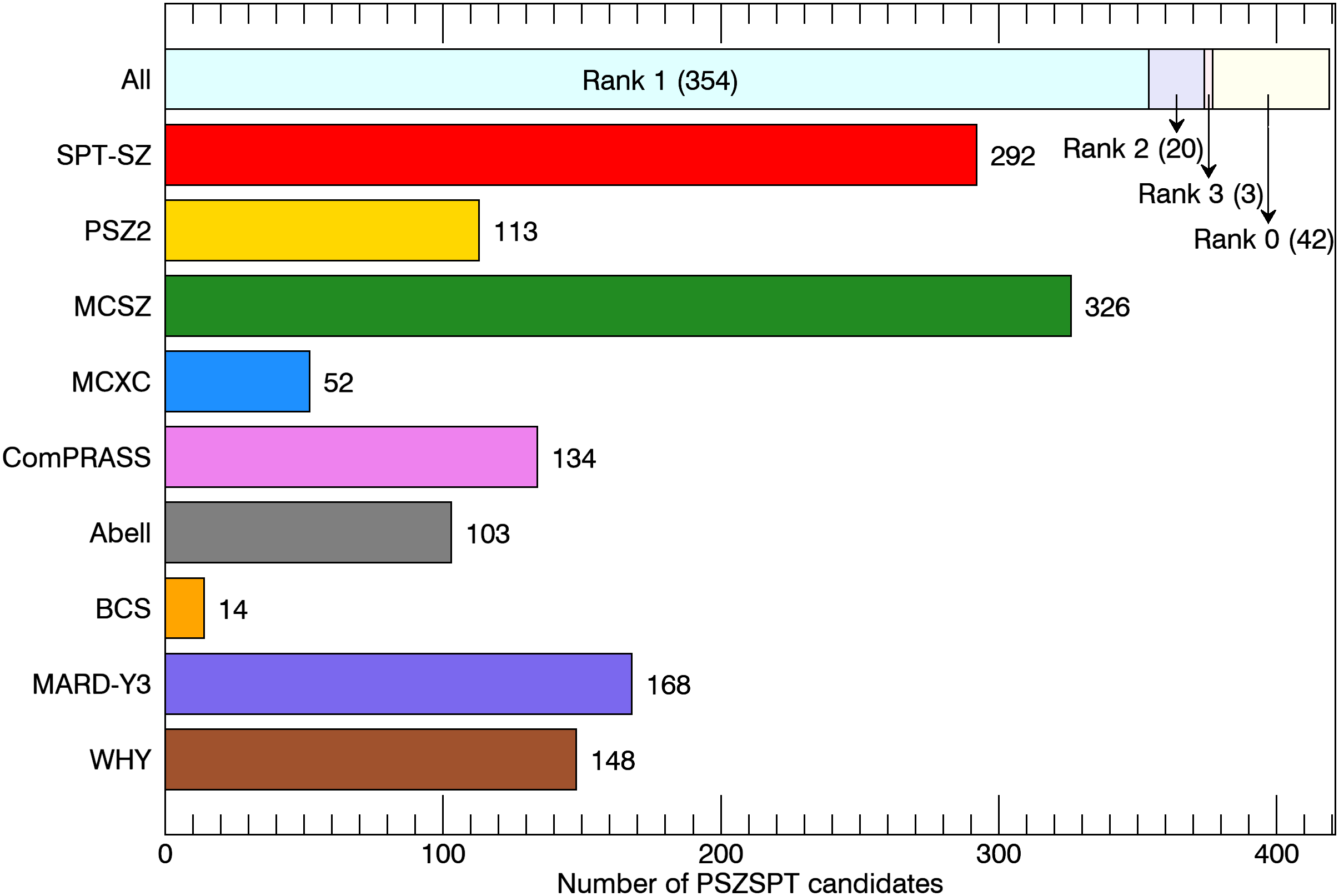}
\caption{{\it First line:} Number of PSZSPT candidates according to their rank. We note that 0=unidentified, 1=associated to known cluster, 2=possibly associated to known cluster, and 3=multiple detection. {\it Other lines:} Number of rank=1 and rank=2 detections associated with external catalogs.}
\label{fig:rank}
\end{figure}

In the first line of Fig.~\ref{fig:rank}, we summarize the number of PSZSPT detections according to their rank. In the other lines, we provide the number of rank=1 and rank=2 detections associated to external catalogs that we considered in this article. We note that the total number exceeds 354+20=374 (numbers given in the first line) because the external catalogs share some common objects.

\subsection{Missed SPT and PSZ2 clusters}
\label{sec:missed}

\begin{figure}
\centering
\includegraphics[width=\hsize]{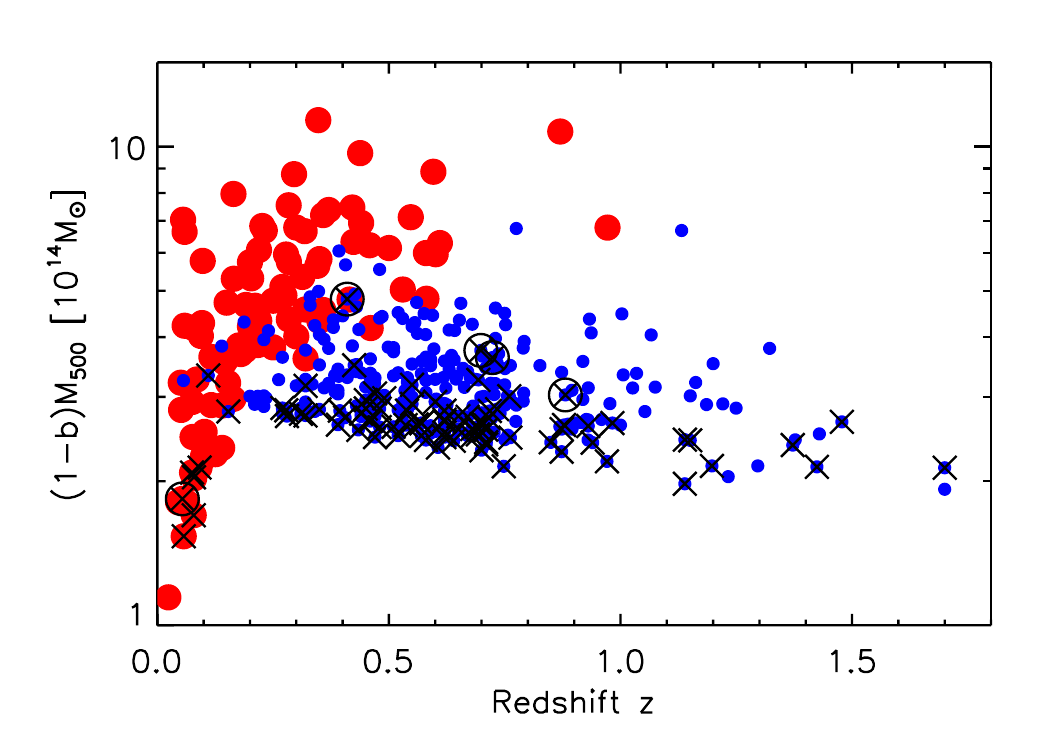}
\caption{Mass-resdhift distribution of the SPT (small blue dots) and PSZ2 (large red dots) clusters. The SPT and \Planck\ clusters that went undetected in the PSZSPT are marked with large black crosses. The large black circles additionally indicate the signal-to-noise greater than seven clusters missed by the PSZSPT. We note that $(1-b)$ is the mass bias factor as in Fig.~\ref{fig:sptsz_match_mass} and \ref{fig:psz2_match_mass}.}
\label{fig:missed}
\end{figure}

We now investigate the SPT and PSZ2 clusters in the SPT-SZ footprint with a signal-to-noise greater than five which are not in the PSZSPT catalog.
The official SPT catalog contains 677/409 detections with a signal-to-noise of $\xi>4.5/5$. After the \cite{chown2018} mask was applied, 379 detections with $\xi>5$ remained. We additionally applied the \Planck\ union mask which leaves 376 $\xi>5$ detections in the SPT catalog. We found 292 matches in Sect.~\ref{sec:identification} between the full SPT catalog ($\xi>4.5$) and the PSZSPT catalog. This reduces to 268 matches with SPT detections at $\xi>5$ in the \cite{chown2018} and \Planck\ union masks. Thus 376-268=108 SPT detections at $\xi>5$ are unmatched with the PSZSPT detections. Seven of them are in the point source mask built by our extraction algorithm (Sect.~\ref{sec:catalogue_construction}). Among the remaining 101, four have $\xi>7$ and 82 have assigned redshift and mass. We plotted the SPT detections having a redshift and mass in Fig.~\ref{fig:missed} as small blue dots. We overplotted the 82 unmatched detections with large black crosses. We added large black circles to the four high $\xi>7$ clusters (they all have a redshift). 

We applied the same methodology to the PSZ2 detections. We note that 107 PSZ2 detections with $S/N>5$ are in the \cite{chown2018} and PSZ2 union masks. Among the 113 matches found in Sect.~\ref{sec:identification}, 91 have $S/N>5$. This leaves 107-91=16 PSZ2 detections unmatched. Five of them are in the point source mask built by our extraction algorithm (Sect.~\ref{sec:catalogue_construction}). Among the remaining 11, one have $S/N>7$ and six have an assigned redshift and mass. We plotted the PSZ2 clusters with a redshift and mass as large red dots in Fig.~\ref{fig:missed}. We overplotted the six unmatched detections with large black crosses and the cluster with $S/N>7$ with a large black circle. We note that we applied a 0.8 recalibrating factor to the SPT mass, as in Fig.~\ref{fig:sptsz_match_mass}. We provide the names of the missed SPT and PSZ2 clusters in Appendix~\ref{sec:missed_names}.

When combining SPT-SZ and \Planck, the $S/N$ of the detections is expected to be, on average, greater than the $S/N$ on the individual experiments. However, due to estimation errors, there is a scatter around the expected value, which makes some of the clusters, especially those close to the limit, to down-scatter below the PSZSPT detection threshold. Additionally, we show in Appendix~\ref{sec:sptszphot} that our extracted $S/N$ on SPT-SZ maps is, on average, 10\% lower than the S/N $\xi$ published by the SPT collaboration. For these reasons, we expected to miss several SPT-SZ clusters close to $\xi=5$, and also a small number of higher signal-to-noise clusters.

The majority of the unmatched candidates (indicated by the black crosses) are indeed at the detection limit of the SPT and \Planck\ catalogs. Thus they may have been missed because of noise fluctuations in the filtered maps. There are, however, four SPT detections and one \Planck\ detection (large black circles) with a signal-to-noise greater than seven. Three of the SPT clusters have $S/N<5$ in the filtered maps after the first pass and are thus not detected by the joint algorithm. The last SPT cluster has $S/N\sim8.1$ in the filtered maps after the first pass. It is thus detected in the first pass of the algorithm, but it was not included in the catalog after the second pass because it is located close to the edge of the SPT-SZ footprint and the noise power spectrum was not estimated properly after the cluster re-centering so the detection is rejected. The PSZ2 undetected cluster has $S/N\sim5.5$ after the first pass of the algorithm, but it does not pass the threshold after the re-centering of the second pass ($S/N\sim4.6$) and is thus not included in the catalog.

\subsection{Completeness}
\label{sec:completeness}

In this section, we adopt the \cite{planck2018cmb} CMB cosmology (TT,TE,EE+lowE+lensing model\footnote{flat model with $h=0.6736$, $\Om=0.3153$, and $\sig=0.8111$}).

We followed the method from Sect. 3.2 of~\cite{planckcosmo2014} (also described in Sect. 3.2 of~\cite{PSZ1}, Sect. 3.3 of~\cite{planckcosmo2016}, and Sect. 4.2 of~\cite{PSZ2}) to predict the completeness of the joint and individual surveys semi-analytically. We assumed that the noise of the maps is Gaussian after filtering with the MMFs. Thus, the completeness can be expressed as an {\sl erf} function of the cluster size $\theta_{500}$, the cluster flux $Y_{500}$, the detection threshold $q$ (set to 5 in this work), and the position on the sky (ra,dec). We express the completeness as a function of redshift $z$ and mass $M_{500}$, adopting the $Y_{500}-M_{500}$ and $\theta_{500}-M_{500}$ relations from Eq.~7 and Eq.~9 in~\cite{planckcosmo2014}. We then integrated the result over the sky coverage of SPT-SZ. 

\begin{figure*}
\centering
\includegraphics[width=0.45\hsize]{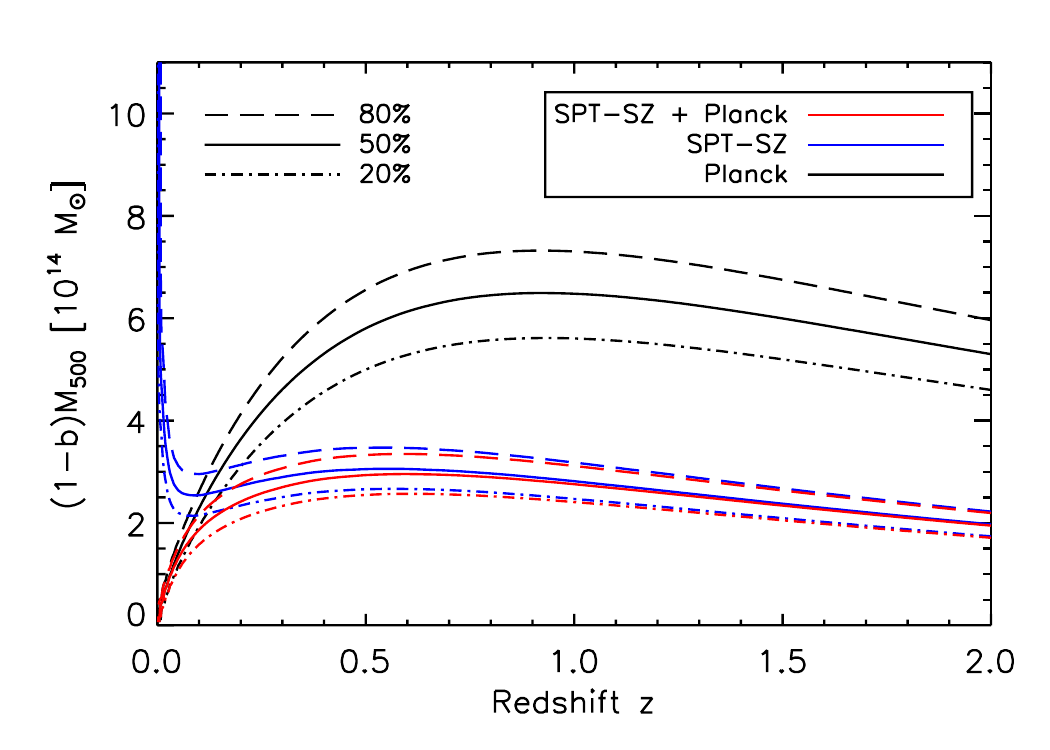}  \includegraphics[width=0.45\hsize]{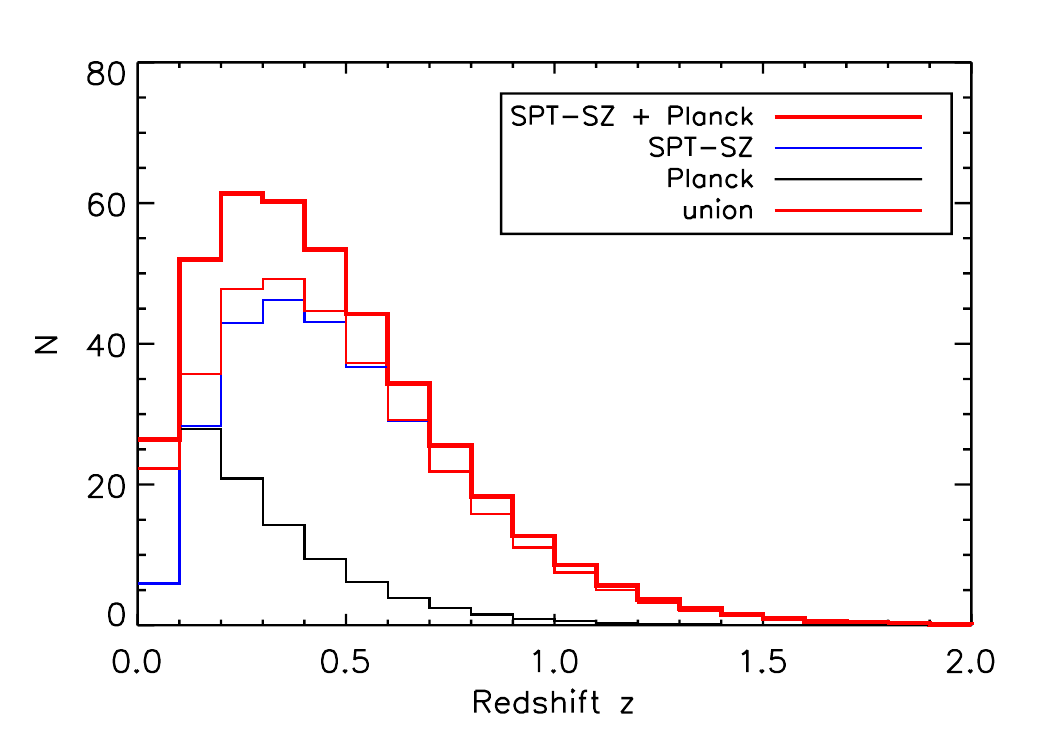}
\caption{{\it Left:} Completeness of the joint \Planck+SPT-SZ survey (red) compared to the completeness of individual \Planck\ (black) and SPT-SZ (blue) surveys. The dashed, solid, and dotted-dashed line correspond to the 80, 50, and 20\% completeness level, respectively. {\it Right:} Predicted cluster counts in each redshift bin for a \cite{planck2018cmb} primary CMB cosmology in the SPT-SZ footprint. The \Planck+SPT-SZ curve corresponds to the joint extraction, while the union curve corresponds to the union of the individual \Planck\ and SPT-SZ catalogs. The mass bias factor $(1-b)$ was set to 0.63 in the right figure.}
 \label{fig:completeness_and_counts}
\end{figure*}

The resulting completeness is shown in the left panel of Fig.~\ref{fig:completeness_and_counts} for the three catalogs: \Planck+SPT-SZ, \Planck, and SPT-SZ. The completeness of the joint catalog \Planck+SPT-SZ is driven by SPT-SZ at redshift $z>0.5$ and by \Planck\ at $z<0.1$. This is expected because SPT-SZ has less instrumental noise and a better resolution than \Planck, leading to better efficiency at detecting high-z clusters. On the other hand, the SPT filtering strategy smoothes the large angular scales, which prevents the detection of very low-z clusters. \Planck, as a satellite, is not affected by this effect and can detect the low redshift clusters. Both surveys contribute to the intermediate redshift range $0.1<z<0.5$. 

We expect the unidentified detections to mainly populate this specific range. We note that $(1-b)$ is the "mass bias factor," which relates the true mass $M_{500}$ to the {\em XMM-Newton}-like mass $M_{500,X}=(1-b)M_{500}$. The scaling laws from Eq.~7 and Eq.~9 in~\cite{planckcosmo2014} are based on {\em XMM-Newton} masses, and we decided to leave this parameter free in this left panel to aid comparison with other works.

From the completeness, we can predict the expected cluster counts from a given mass function. We chose the \cite{tinker2008} mass function, and we fixed (1-b)=0.63 to adjust the \Planck\ MMF3 cosmological cluster counts to the  \cite{planck2018cmb} CMB cosmology. This value of $(1-b)$ is very close to the value $(1-b)=0.62 \pm 0.03$ obtained by~\cite{planck2018cmb} (see their Eq.~34) using the same dataset, and to the value $(1-b)=0.622 \pm 0.033$ found by~\cite{salvati2019} (see their Table~1) on an extended dataset. The predicted redshift distribution is given in the right panel of Fig.~\ref{fig:completeness_and_counts} for the SPT-SZ footprint. 

As expected from the completeness, the overall \Planck+SPT-SZ cluster count (thick red line) is dominated by the SPT-SZ dataset, except at a very low redshift where \Planck\ provides the information. The predicted total number of clusters is 413, 302, and 111 for \Planck+SPT-SZ, SPT-SZ, and \Planck,\ respectively, which is in very good agreement with the number of detections given in Sect.~\ref{sec:catalogue_construction}. The union catalog of the individual \Planck\ and SPT-SZ catalogs is shown as the thin red line for comparison with the joint catalog. We provide these predicted cluster numbers as a consistency check between the cluster counts of our joint catalog and the \cite{planck2018cmb} CMB cosmology. Constraining cosmological parameters from this joint catalog is beyond the scope of this work.

\section{Summary, discussion, and future work}
\label{sec:sum}

We performed, for the first time, an SZ cluster extraction using space-based (\Planck) and ground-based (SPT) data jointly.\ We also provide the associated PSZSPT catalog of 419 sources at $S/N>5$.

For this purpose, we modified the MMF3 algorithm that was initially developed for \Planck\ data to make it compatible with ground based data. The main difficulties were including the transfer function of the SPT survey as well as handling the high resolution ground-based and low resolution space data at the same time, which required the use of small (0.43 arcmin) pixels on large ($10 \times 10 \deg^2$) tangential maps.
In the process of building the joint catalog, we thoroughly characterized the SPT public maps (transfer function, point source, and cluster photometry) with respect to the SPT official catalogs (Sect.~\ref{sec:public_consistency}, Appendix~\ref{sec:filterfunc} and~\ref{sec:sptszphot}). We found good agreement between quantities extracted with our tools and the official SPT catalogs. However, the S/N of our SPT extractions is 10\% lower than the published values. We attribute this to the difference in map depths between the SPT public maps~\citep{chown2018} and the maps used in~\cite{bleem2015}. Our extraction method would thus be more efficient if it could be applied to the maps used in~\cite{bleem2015}. We also checked that our new extraction method provides results consistent with the \Planck\ publications (Sect.~\ref{sec:public_consistency}, Appendix~\ref{sec:plckphot}). We then showed that the \Planck\ and SPT data provide consistent flux measurements for SPT and \Planck\ clusters, respectively (Sect.~\ref{sec:sptplanck_consistency}).

We cross-matched the PSZSPT catalog with other cluster catalogs in the SPT-SZ footprint. We checked for remaining unidentified detections in SIMBAD and NED. We note that 292(113) detections are matched or possibly matched to SPT-SZ(PSZ2) detections, respectively, with 82 being common to both of the catalogs. We identified 47 detections with clusters in catalogs other than SPT-SZ and PSZ2, and seven additional detections using SIMBAD and NED. Finally, we could not identify counterparts for the remaining 42 detections which need to be validated by future external follow-ups (Sect.~\ref{sec:identification}).

We finally estimated the completeness of the PSZSPT catalog and checked that the extracted counts are consistent with the standard $\Lambda {\rm CDM}$ model when adopting the \Planck\ cluster modeling and scaling laws (Sect.~\ref{sec:completeness}).
The PSZSPT catalog is described in Appendix~\ref{sec:description} and a complete version is available in electronic format.

The increase in the number counts from the joint catalog with respect to the union of the space- and ground-based catalog is moderate, as shown in Fig.~\ref{fig:completeness_and_counts} (difference between the red thin and thick lines). The majority of new detections are expected to be in the redshift range [0.1,0.6] around the location where the \Planck\ and SPT-SZ completenesses cross. The most interesting application of the space- and ground-based joint cluster analyses with current data sets may thus be astrophysical studies and, in particular, cluster profiles, the space-based (ground-based) data providing the large-scale (small-scale) information, respectively. We leave this work for a future article.

This first proof of concept of joint cluster extraction with space- and ground-based data opens the path to other possible catalogs when new data will be publicly available, for example \Planck\ and ACT~\citep{naess2020} or \Planck\ and SPT-ECS~\citep{bleem2020}.
But the approach would be most useful in the case of the longer term ground-based Simons Observatory~\citep{simonsobservatory} or CMB-S4~\citep{cmbs4} and proposed space mission such as {\em PICO}~\citep{hanany2019} or {\em Backlight}~\citep{basu2019,delabrouille2019}, which have resolutions matching between space and ground (1~arcmin FWHM at 300~GHz) and together cover a large frequency range from a few tens of gigahertz to terahertz. The space and ground approach will be crucial to disentangle the various emissions in clusters:  the different SZ effects as well as the radio and infrared emission of galaxies hosted by clusters.

\begin{acknowledgements}
We would like to thank the anonymous referee for useful comments which helped to clarify this work. We acknowledge the use of the Legacy Archive for Microwave Background Data Analysis (LAMBDA), part of the High Energy Astrophysics Science Archive Center (HEASARC). HEASARC/LAMBDA is a service of the Astrophysics Science Division at the NASA Goddard Space Flight Center. We also used the HEALPix software~\citep{gorski2005} available at \url{https://healpix.sourceforge.io}  and the WebPlotDigitizer by Ankit Rohatgi.
This research has also made use of the M2C Galaxy Cluster Database, constructed as part of the ERC project M2C (The Most Massive Clusters across cosmic time, ERC-Adv grant No. 340519), the SIMBAD database operated at CDS, Strasbourg, France~\citep{wenger2000}, and the NASA/IPAC Extragalactic Database (NED), which is funded by the National Aeronautics and Space Administration and operated by the California Institute of Technology.
J.-B. Melin thanks Monique Arnaud for constructive discussions which helped to improve this article and for suggesting the acronym PSZSPT. He also warmly thanks the high school students Julie Hebert, Aubin Mayer, Hugo Levandowski and Ana{\"e}lle Meurant who helped him search the PSZSPT locations in the SIMBAD and NED databases.
A portion of the research described in this paper was carried out at the Jet Propulsion Laboratory, California Institute of Technology, under a contract with the National Aeronautics and Space Administration.
\end{acknowledgements}

\appendix

\section{Characterization of the SPT-SZ filter transfer function}
\label{sec:filterfunc}

\begin{figure*}
\centering
\includegraphics[width=0.45\hsize]{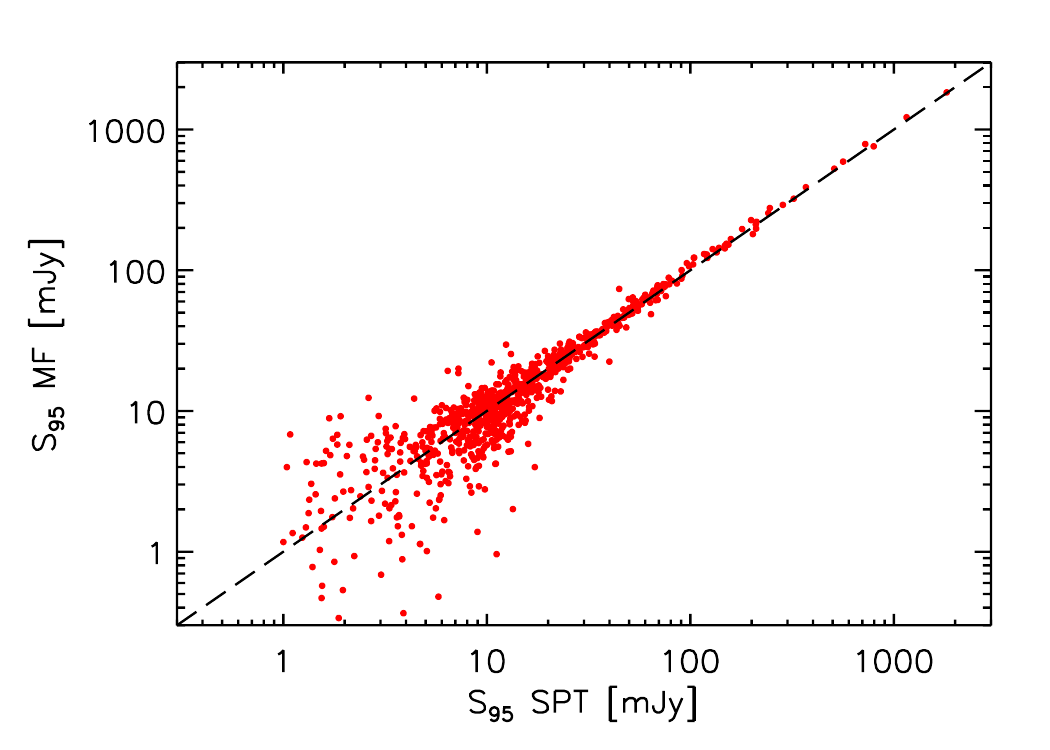}  \includegraphics[width=0.45\hsize]{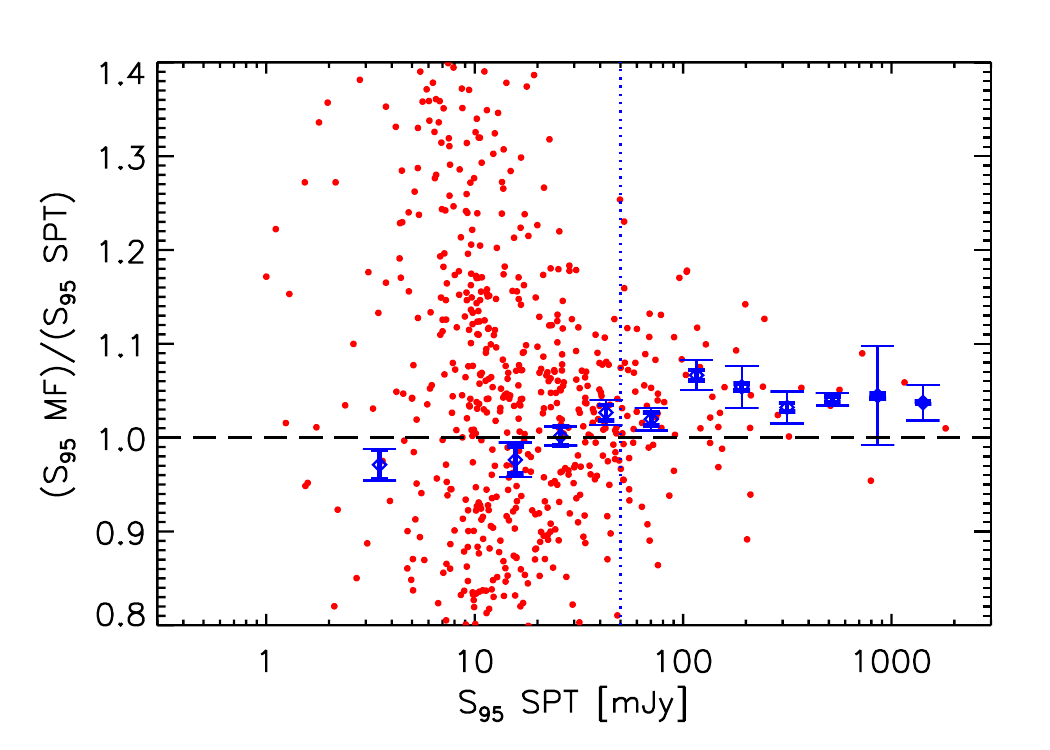}
\includegraphics[width=0.45\hsize]{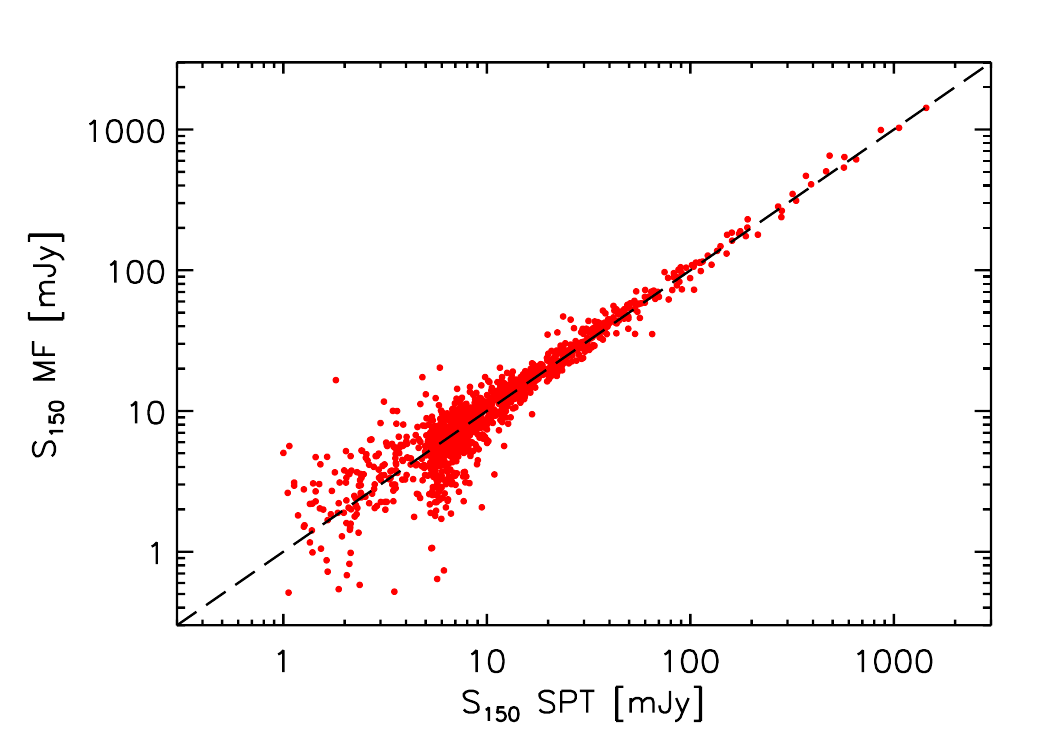}  \includegraphics[width=0.45\hsize]{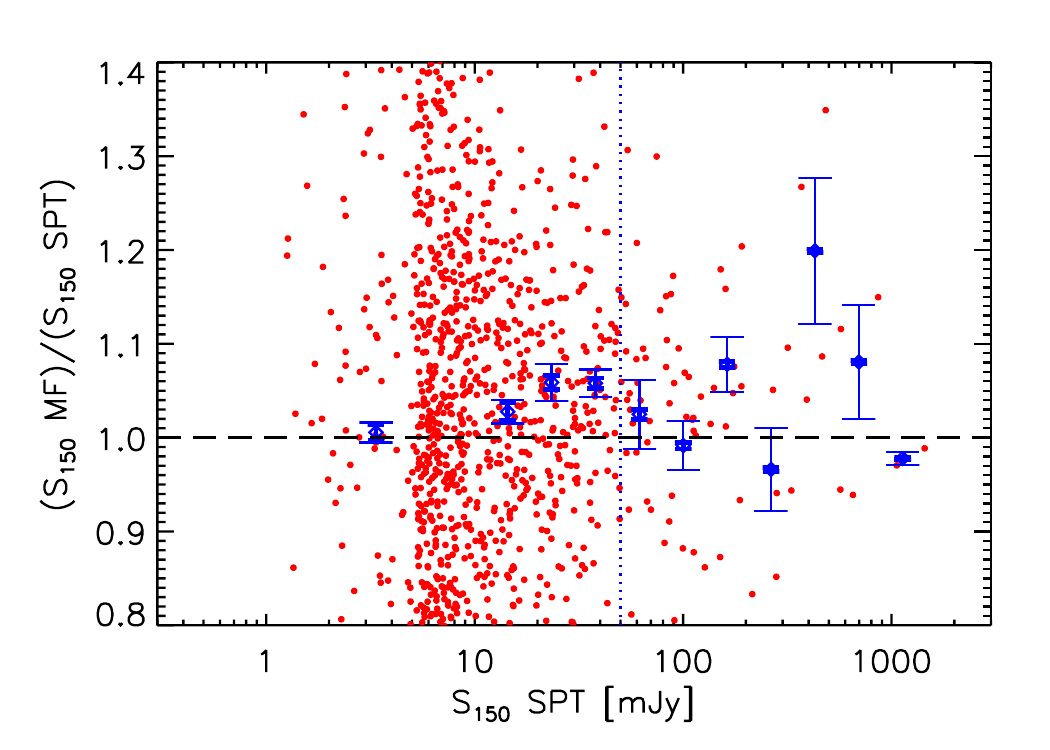}
\includegraphics[width=0.45\hsize]{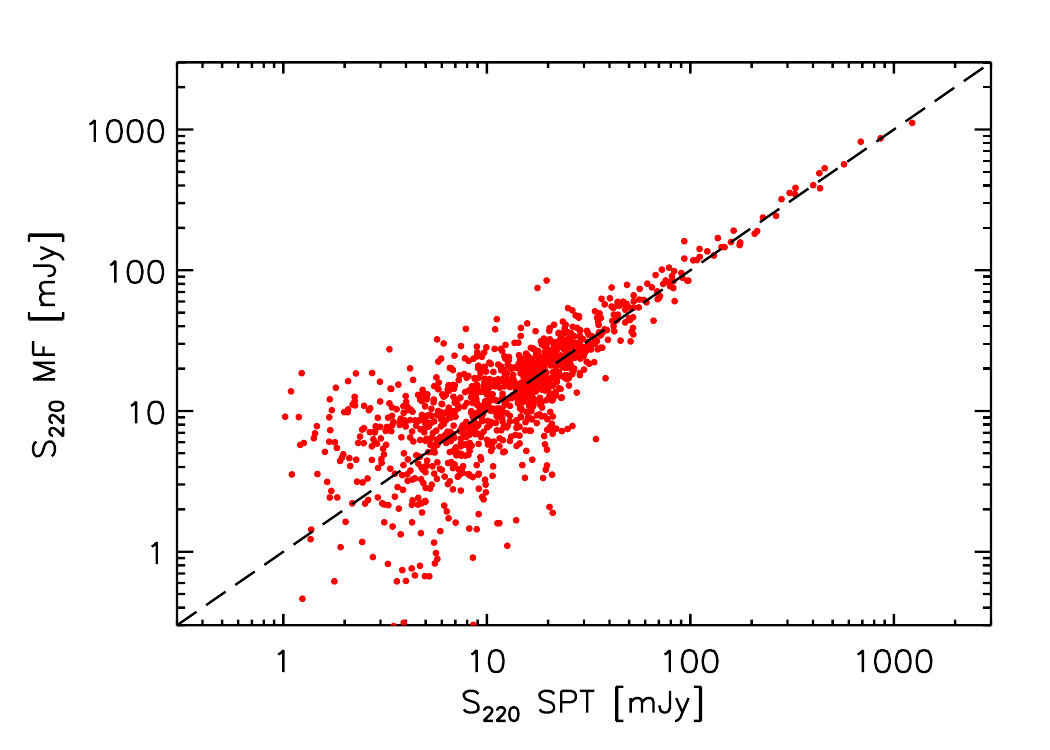}  \includegraphics[width=0.45\hsize]{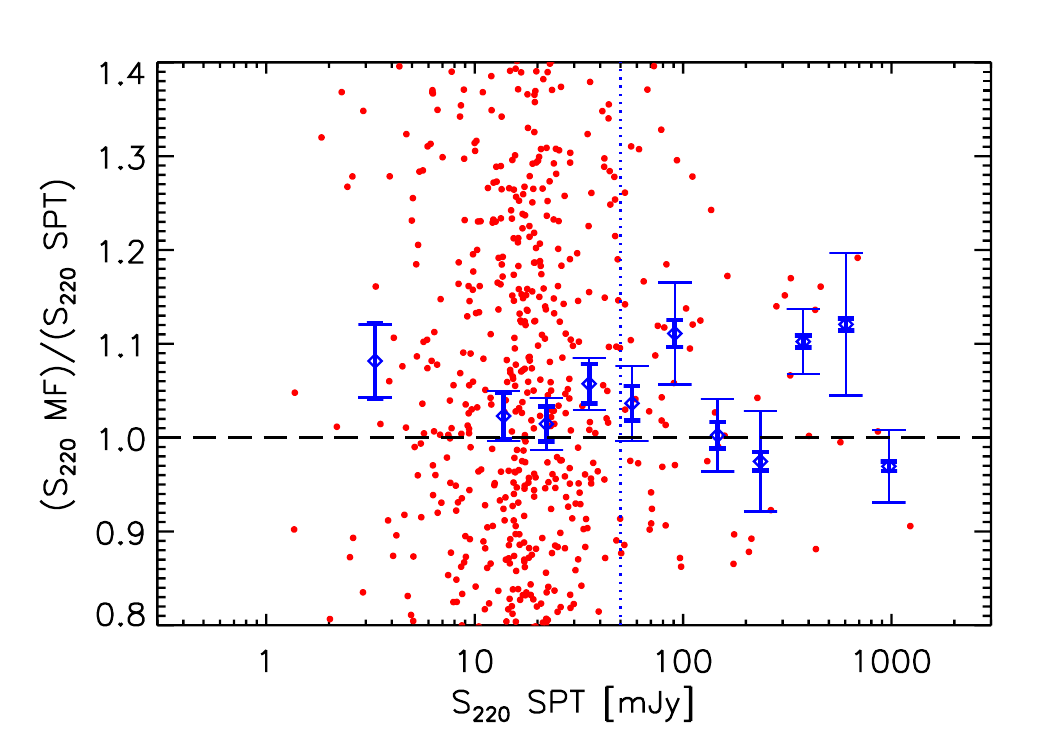}
\caption{{\it Left column:} Our single frequency matched filter flux versus published SPT flux for the point sources published in~\cite{mocanu2013} from 95~GHz (top) to 220~GHz (bottom). Red dots are individual point sources. There is overall agreement between our recovered fluxes and the values published by the SPT collaboration. {\it Right column:} Zoom-in on the ratio between the two flux measurements as a function of the SPT flux. Blue diamonds are weighted averages. Thick bars display 68\% statistical errors, and thin bars show 68\% errors obtained by the bootstrap method. Despite the global agreement in the log-log plane shown in the left column, the ratio is significantly greater than unity, in particular at a large flux (S>50~mJy) for the three SPT frequencies. We indicate this 50~mJy limit by the vertical blue dotted lines in the three panels.}
              \label{fig:psfluxmf_vs_psfluxspt}
\end{figure*}

The SPT-SZ filter transfer function is provided with the public data for each SPT frequency. We tested it by comparing point source fluxes extracted from the public data (using the provided transfer function) and the fluxes published for the same sources by the SPT collaboration.  In practice, we adopted the positions of the point sources in the 2013 catalog~\citep{mocanu2013} and extracted their flux using a single frequency matched filter applied to the public data. We used the SPT-SZ frequency responses given in Fig.~10 of~\cite{chown2018} (long-dashed lines). The results are presented in Figure~\ref{fig:psfluxmf_vs_psfluxspt} for the three SPT channels. 

We restricted the extraction to sources with a published SPT flux above 1~mJy. The left-hand column shows our extracted flux as a function of the published SPT flux for the 95 (top figure), 150 (middle), and 220 (bottom)~GHz channels. There is global agreement between the two. The right-hand column shows the ratio of the two fluxes versus the SPT flux. Blue diamonds are weighted averages of the individual measurements (red dots).  We see that our fluxes are, in fact, systematically overestimated, in particular for flux S above 50~mJy (indicated by the vertical blue dotted line) for the three SPT frequencies. The overestimation is about 5\% with respect to the published values. For fluxes below 50~mJy, there is no significant overestimation in the 95~GHz channel, but there remains an overestimation for the 150 and 220~GHz channels, although less significant than for point sources with fluxes above 50~mJy.

\begin{figure*}
\centering
\includegraphics[width=\hsize]{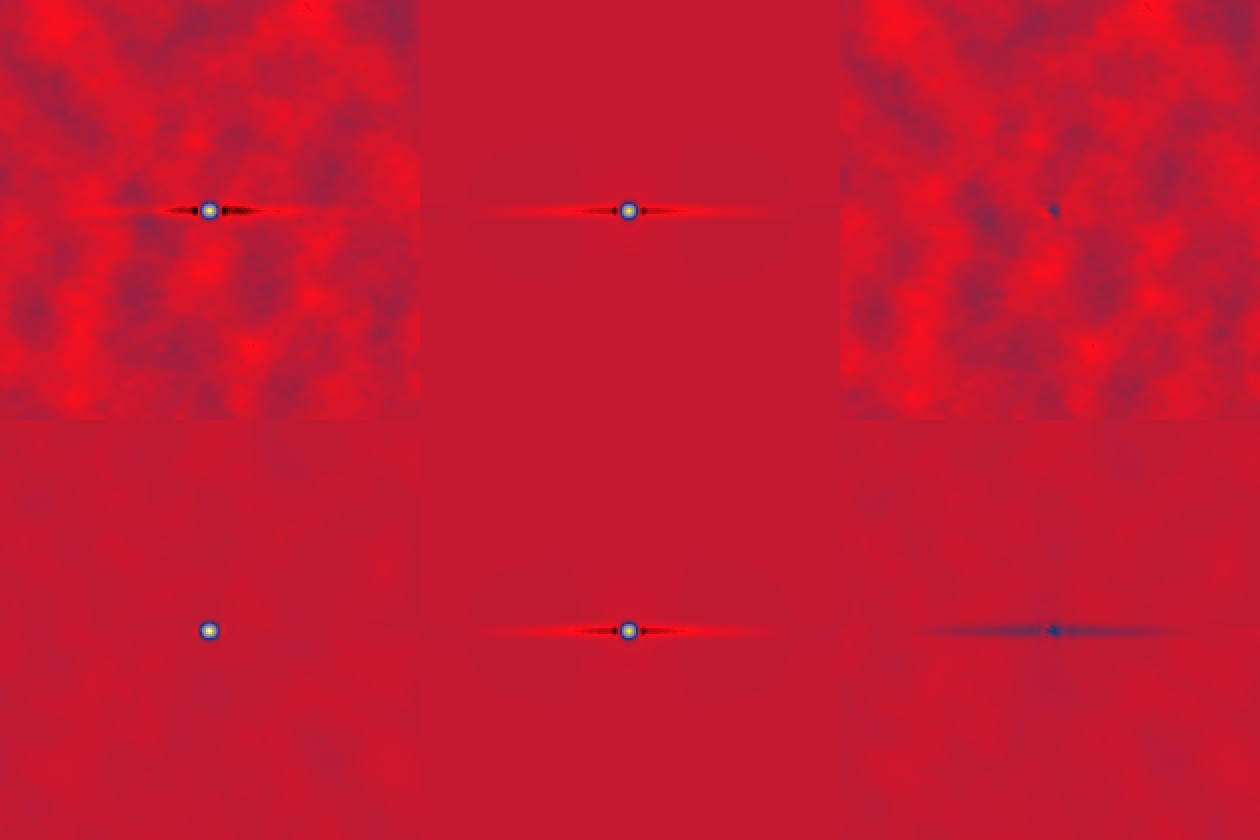}
\caption{Stacks of the 150~GHz point sources of~\cite{mocanu2013} from the public SPT-SZ data in equatorial coordinates (a horizontal line lies at constant declination). The six square panels are 1.5~deg on a side. {\it Top left:} Stack of point sources with published SPT flux below 50~mJy, normalized to the total flux of the stack computed from SPT published values. {\it Top middle:} Gaussian beam (FWHM=1.75~arcmin) convolved with the filter transfer function. {\it Top right:} Difference between top left and top middle images. The residuals are small. {\it Bottom left:} Stack of point sources with published SPT flux above 50~mJy, normalized to the total flux of the stack computed from published SPT  values. {\it Bottom middle:} Identical to top middle, i.e., Gaussian beam (FWHM=1.75~arcmin) convolved with the filter transfer function. {\it Bottom right:} Difference between bottom left and bottom middle images. The residuals are significant. In particular, the stack does not show the negative tails before and after the central maximum in right ascension, which can be seen in black in the middle image. This difference in the patterns can be seen in the difference image as the horizontal trail.}
              \label{fig:psstacks}
\end{figure*}

To investigate the origin of this effect, we stacked the SPT-SZ maps at the locations of the bright (S>50~mJy) and faint (S<50~mJy) point sources separately. We then compared the two stacks to the beam convolved by the filter transfer function.  Figure~\ref{fig:psstacks} presents the results for the 150~GHz sources. The left-hand column gives the two stacks (S<50~mJy top, S>50~mJy bottom). The middle column shows the beam convolved by the filter transfer function (duplicated top and bottom). The right-hand column shows the difference between the first two columns. The stack of the faint sources displays the same pattern as the beam convolved with the filter transfer function, but the stack of the bright sources does not. This result confirms that the beam convolved by the filter transfer function does not correctly model the shape of the bright (S>50~mJy) point sources in the data.  A similar analysis of the 95~GHz and 220~GHz channels leads to the same conclusion, with different patterns in the stacks of point sources with a flux above and below 50~mJy: The stacks of sources with fluxes below 50~mJy show a pattern identical to the beams convolved by the filter transfer functions, while the stacks of sources with fluxes above 50~mJy do not. This result is expected because the regions surrounding the bright (S>50~mJy) point sources were excluded from the fitting of the time stream data~\citep[see Sect. 4.1.1 of][]{chown2018}, so the filter transfer functions are not expected to model these regions of the maps.

The fact that bright sources are not well modeled by the beam convolved by the filter transfer function is the most probable explanation for the overestimation of the SZ flux of bright clusters ($Y^{0.75}\, {\rm SPT}>2 \times 10^{-4} \, {\rm arcmin}^2$) in the SPT-SZ data with respect to the published SPT values. This issue is described in Appendix~\ref{sec:sptszphot} and illustrated in Fig.~\ref{fig:fluxmmf3_vs_fluxspt}.

\section{SPT-SZ photometry}
\label{sec:sptszphot}

We re-extracted the flux and size of the SPT-SZ clusters, fixing the position to the coordinates provided in the SPT cluster catalog~\citep{bleem2015}. The goal was to check the consistency between our photometry (i.e., flux and size estimation) and the photometry from the SPT collaboration. We adopted the cluster modeling of the SPT collaboration: $\beta$ profile of size $\theta_c$ ranging from 0.25 to 3 {\rm arcmin} in steps of 0.25 and $\beta=1$ fixed. We allowed the filter size to vary in the aforementioned range and, for each cluster, we kept the size that maximizes the signal-to-noise. We then extracted the cluster flux for this specific size.

\begin{figure*}
\centering
\includegraphics[width=0.45\hsize]{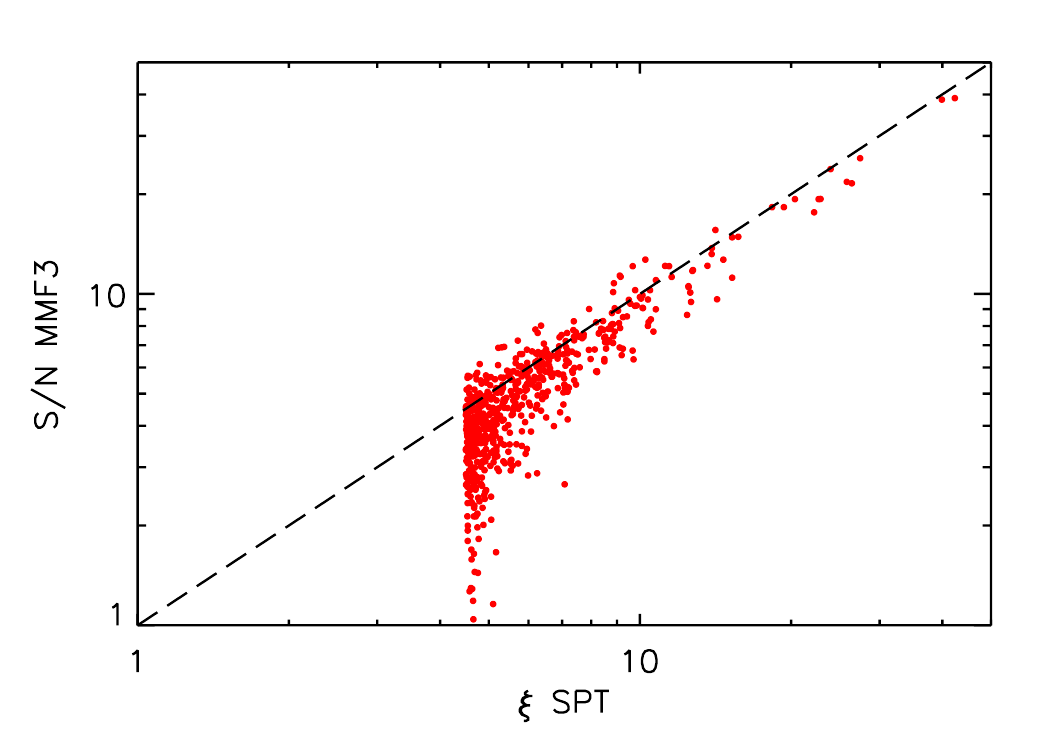}  \includegraphics[width=0.45\hsize]{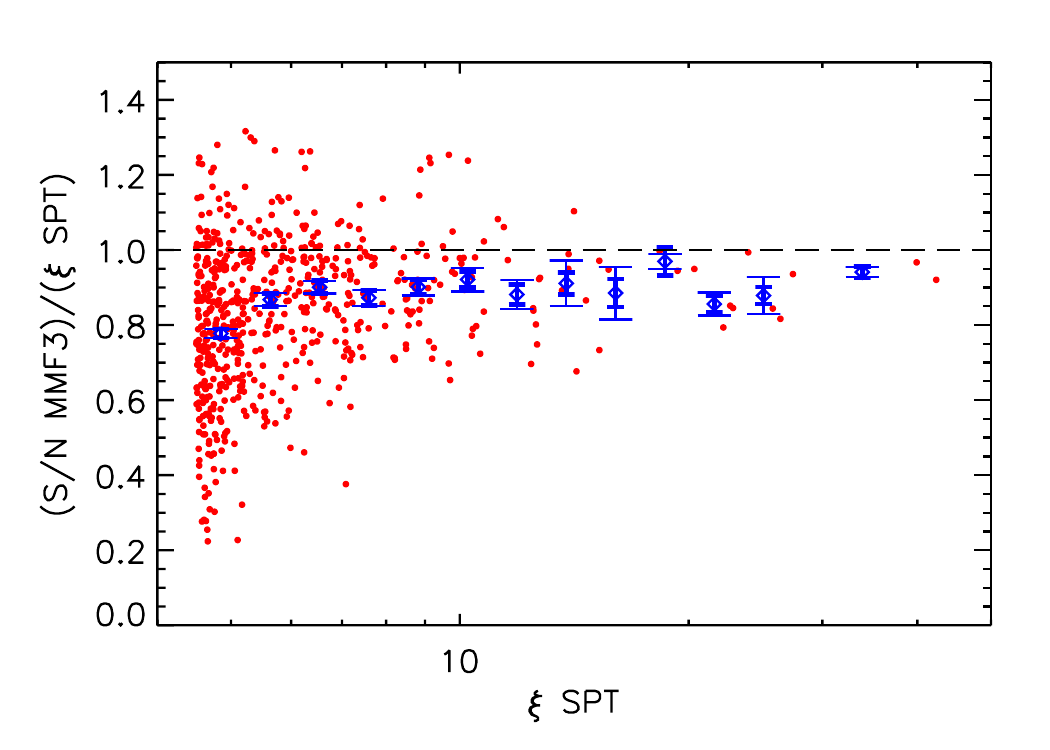}
\caption{{\it Left:} S/N of the SPT clusters extracted from the public SPT data with our modified MMF3 algorithm as a function of the signal-to-noise, $\xi$, published by the SPT collaboration~\citep{bleem2015}. Despite overall agreement, our S/N is systematically lower than the signal-to-noise published by the SPT collaboration. {\it Right:} Ratio of the two signal-to-noise values as a function of the signal-to-noise published by the SPT collaboration. Our S/N is, on average, 0.9 times $\xi$ for $\xi>5$. Red dots are individual clusters. Blue diamonds are weighted averages. Thick error bars display 68\% statistical errors, and thin error bars show 68\% errors obtained by the bootstrap method.}
              \label{fig:snrmmf3_vs_xispt}
\end{figure*}

Figure~\ref{fig:snrmmf3_vs_xispt} compares our extracted signal-to-noise (maximum across filter scales) to the signal-to-noise published by the SPT collaboration, $\xi$. The left-hand panel shows that our S/N is in global agreement with the published SPT signal-to-noise, although it is systematically lower. The right-hand panel shows the ratio of the two signal-to-noise values. Our S/N is, on average, 0.9 times $\xi$ for $\xi>5$.
The SPT public maps~\citep{chown2018} are based on the same 2008-2011 observation cuts as~\cite{mocanu2019}, except for the 150~GHz field based on the observation cuts from~\cite{story2013}. Thus, they are shallower than the maps used in~\cite{bleem2015} (see Table~1 therein). We suspect that the S/N bias to lower values is due to the difference in map depths. We would expect this bias to disappear if the MMFs were applied to the same SPT data as used in~\cite{bleem2015}.

\begin{figure*}
\centering
\includegraphics[width=0.33\hsize]{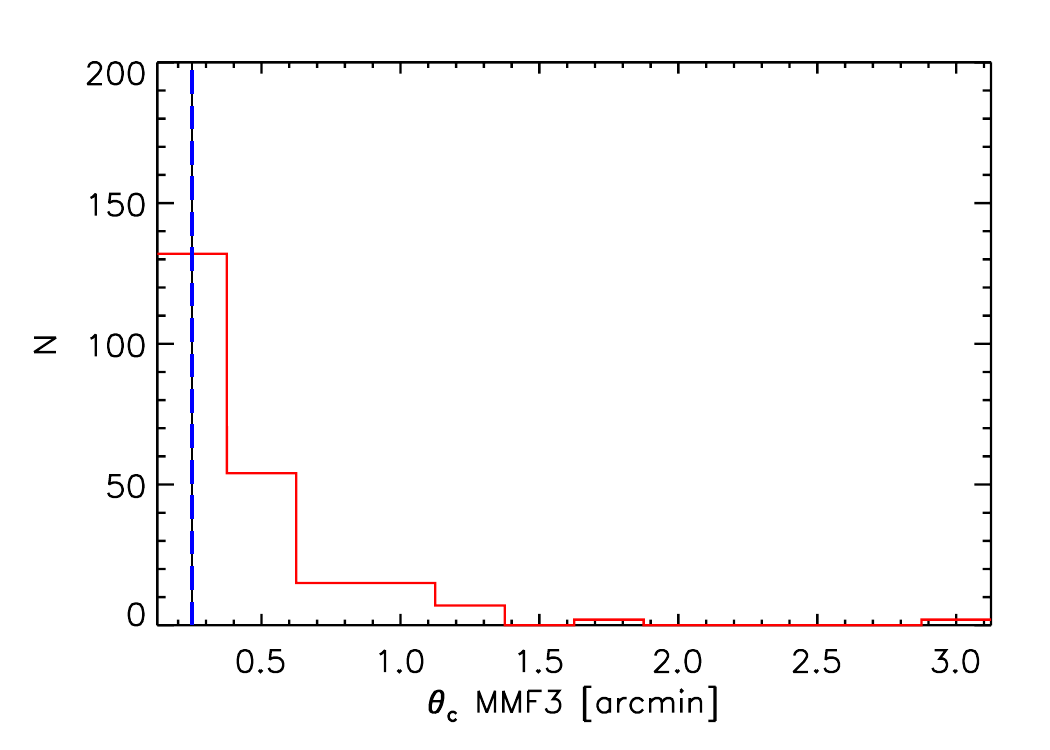}  \includegraphics[width=0.33\hsize]{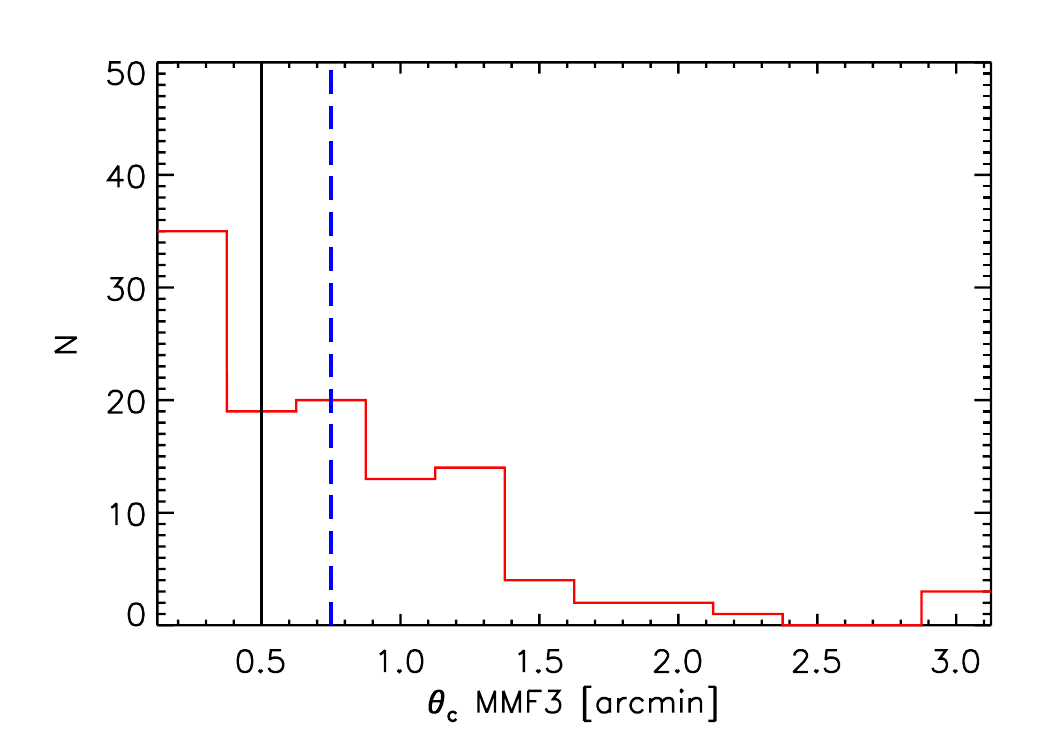} \includegraphics[width=0.33\hsize]{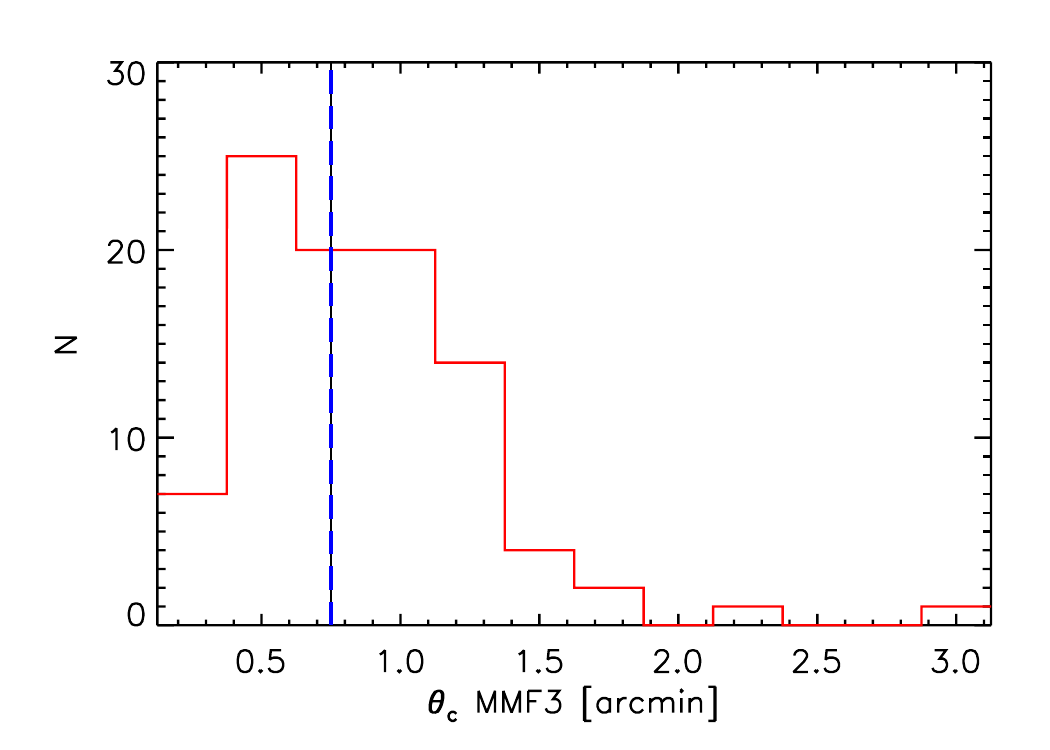}
\includegraphics[width=0.33\hsize]{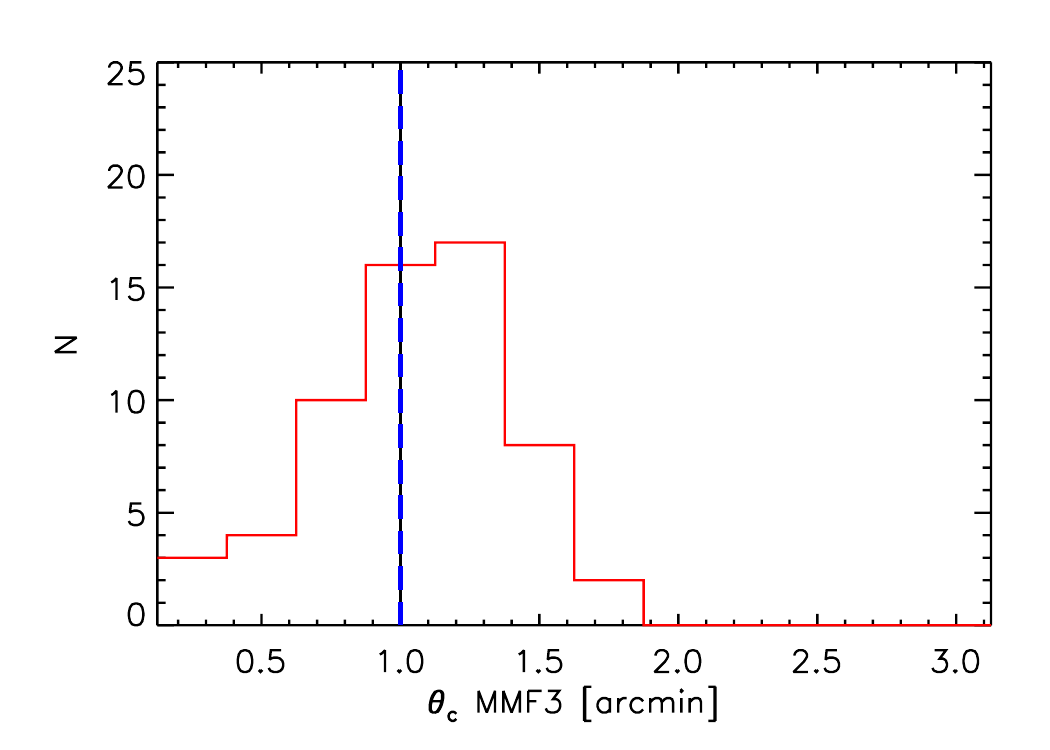}  \includegraphics[width=0.33\hsize]{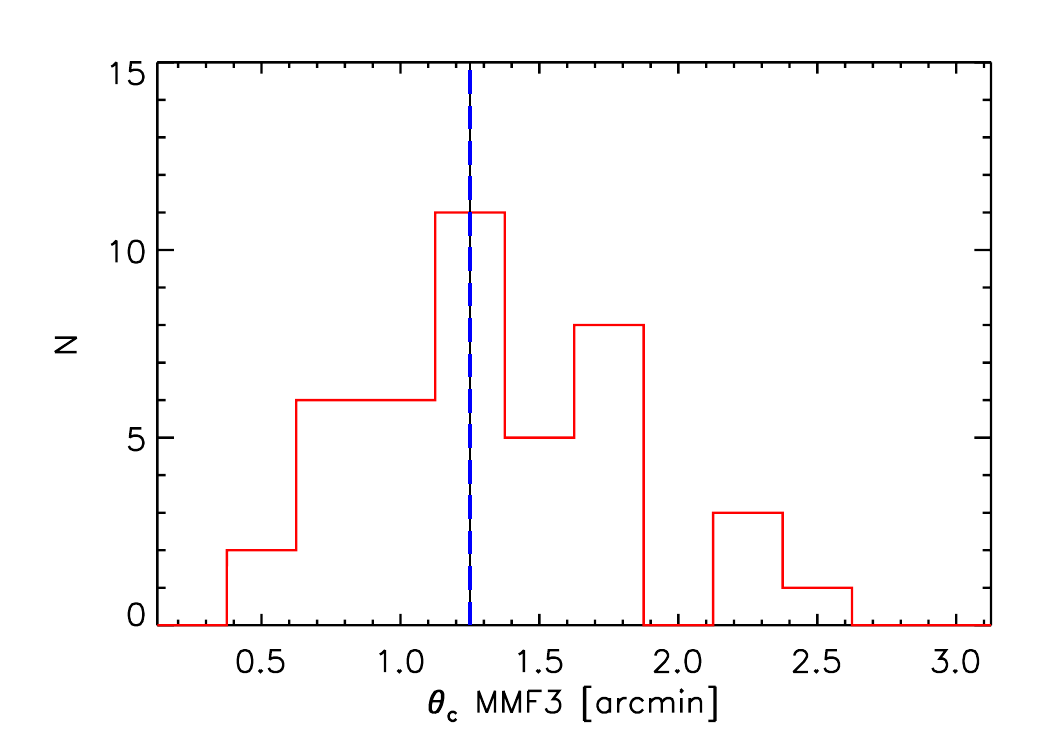} \includegraphics[width=0.33\hsize]{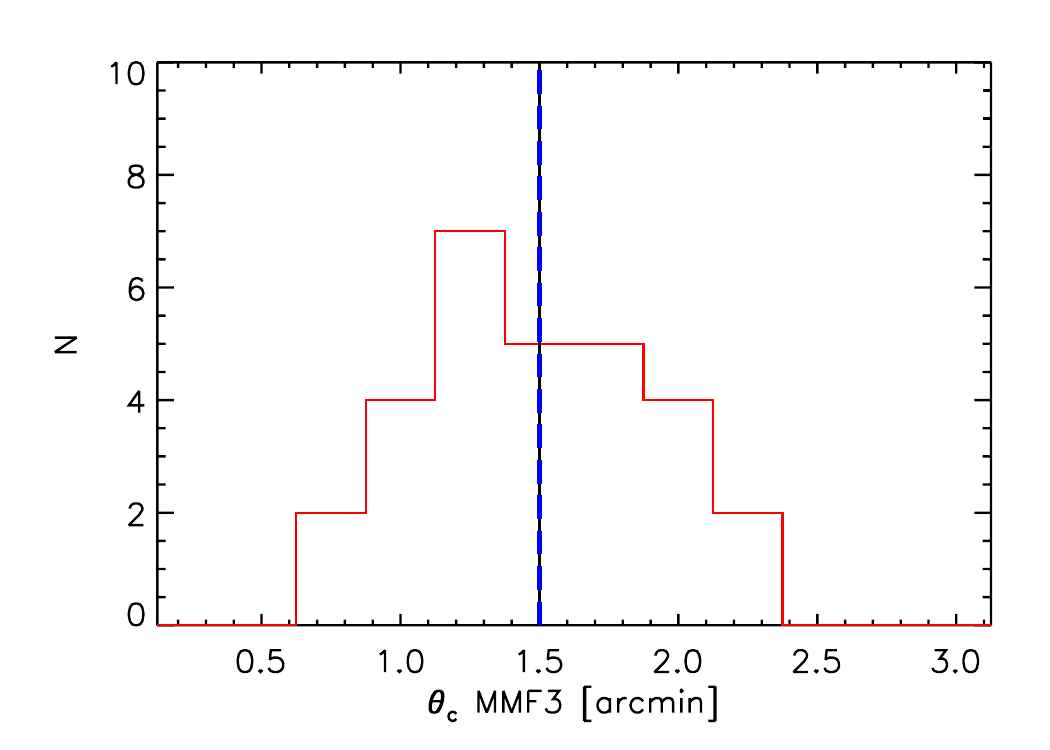}
\includegraphics[width=0.33\hsize]{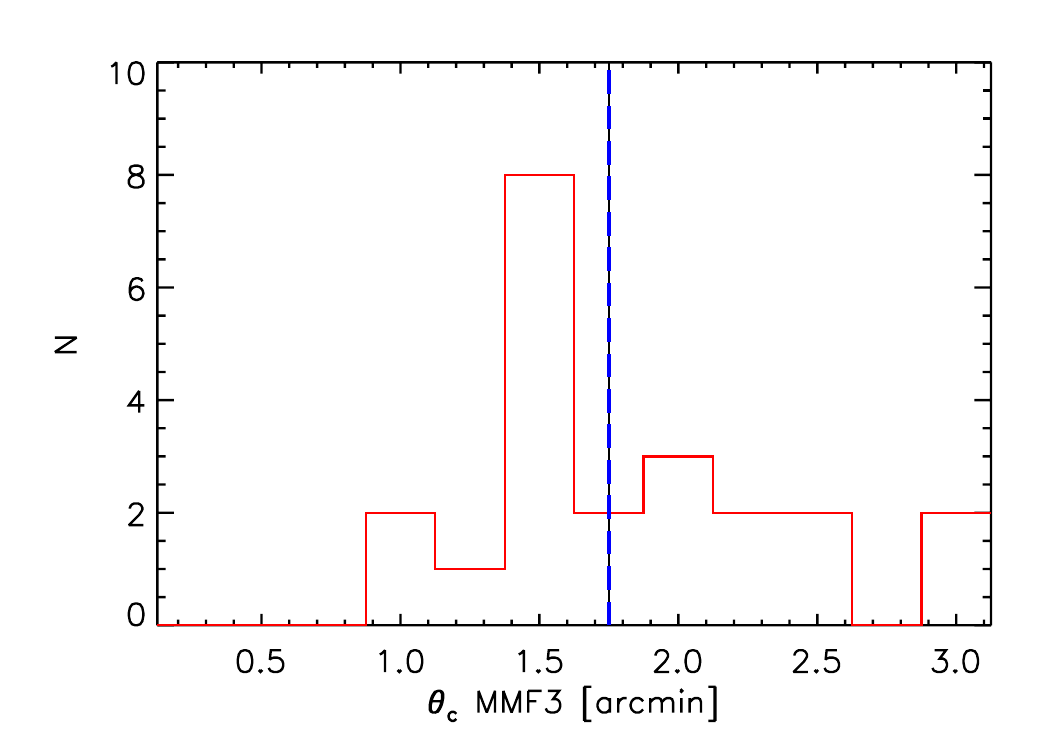}  \includegraphics[width=0.33\hsize]{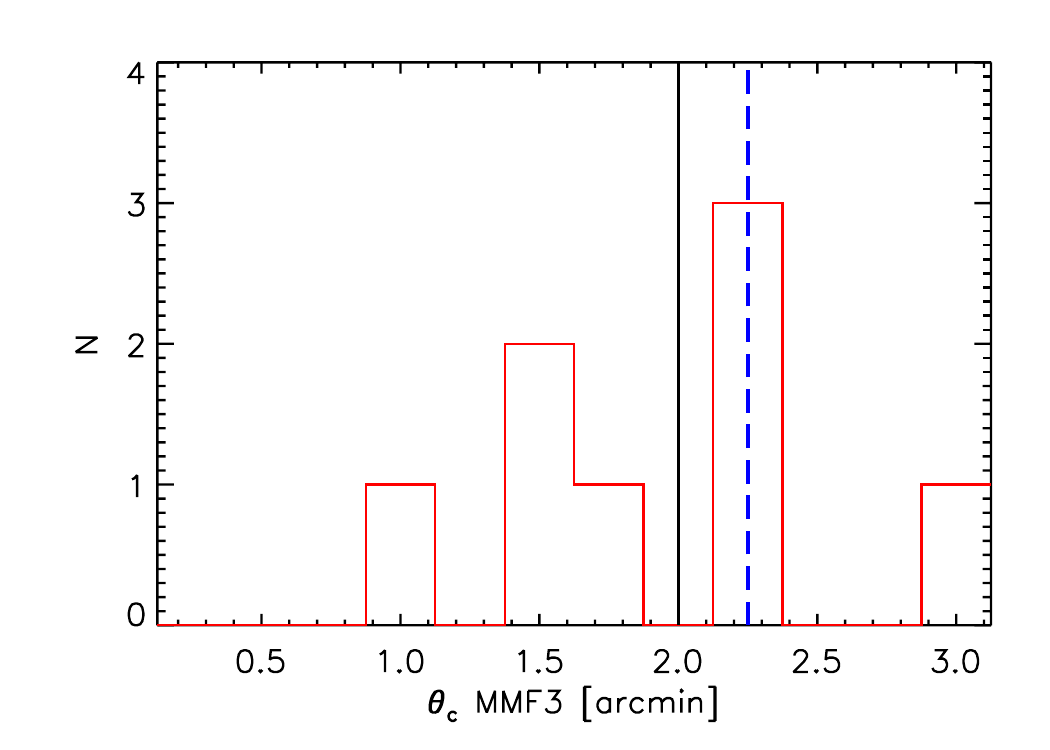} \includegraphics[width=0.33\hsize]{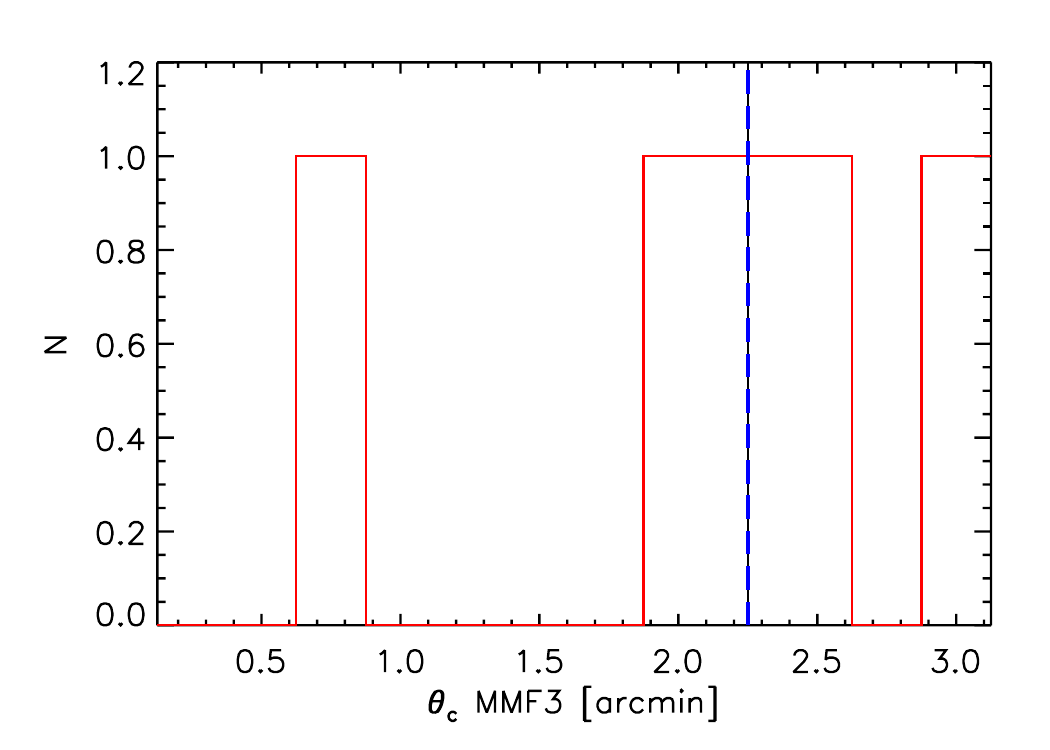}
\includegraphics[width=0.33\hsize]{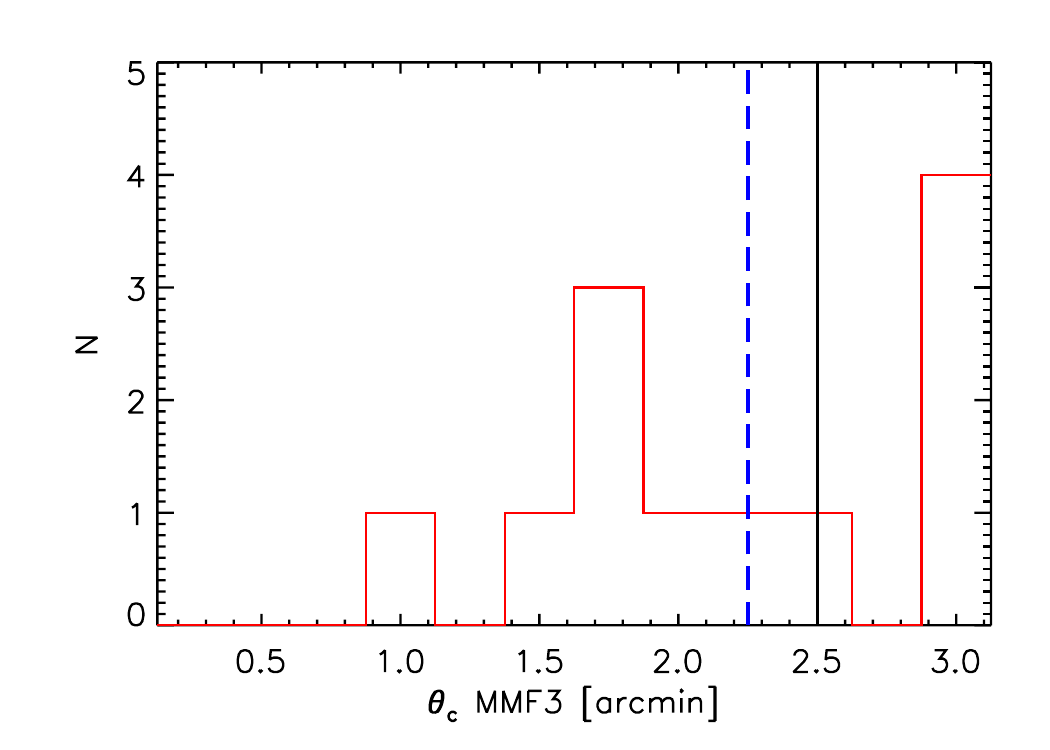}  \includegraphics[width=0.33\hsize]{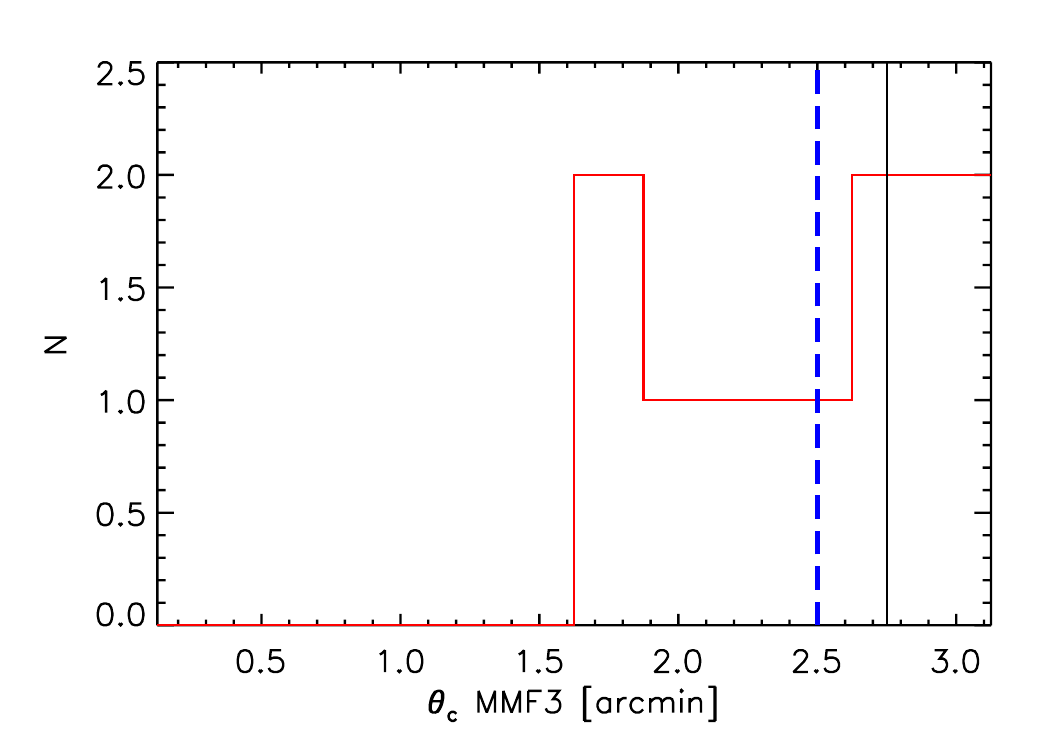} \includegraphics[width=0.33\hsize]{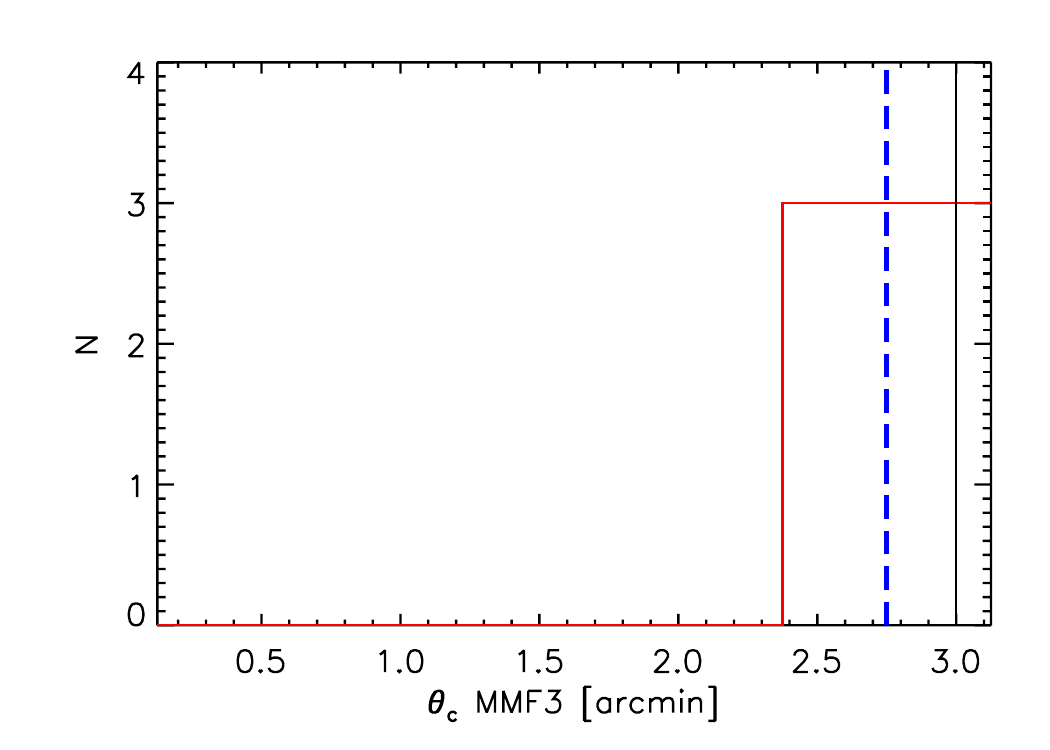}
\caption{Histograms of cluster size as recovered by our algorithm. Each histogram corresponds to clusters with the same size $\theta_c$ determined by the SPT collaboration. The vertical black line displays the size from the SPT collaboration increasing from 0.25 to 3~arcmin in steps of 0.25~arcmin, from left to right and from top to bottom. The red histogram shows the distribution of the size recovered by our algorithm and the thick dashed vertical blue line is the median value of our recovered values. Our recovered sizes are in good agreement with the sizes published by the SPT collaboration.}
              \label{fig:sizemmf3_vs_sizespt}
\end{figure*}

Figure~\ref{fig:sizemmf3_vs_sizespt} compares our recovered blind size to the size published by the SPT collaboration. Each histogram corresponds to clusters with the same size $\theta_c$ determined by the SPT collaboration. Our recovered sizes (red histograms) show no significant deviation from the size published by the SPT collaboration (vertical black line). The thick dashed vertical blue line shows the median of our recovered sizes, which is in good agreement with the SPT size.

\begin{figure*}
\centering
\includegraphics[width=0.45\hsize]{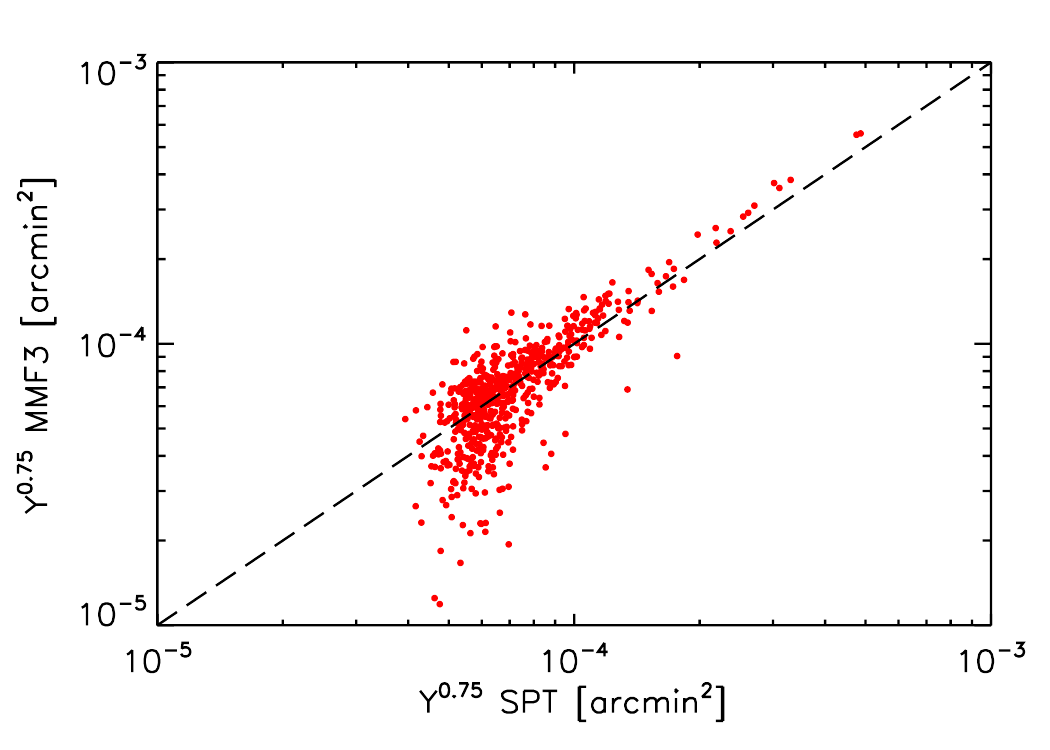}  \includegraphics[width=0.45\hsize]{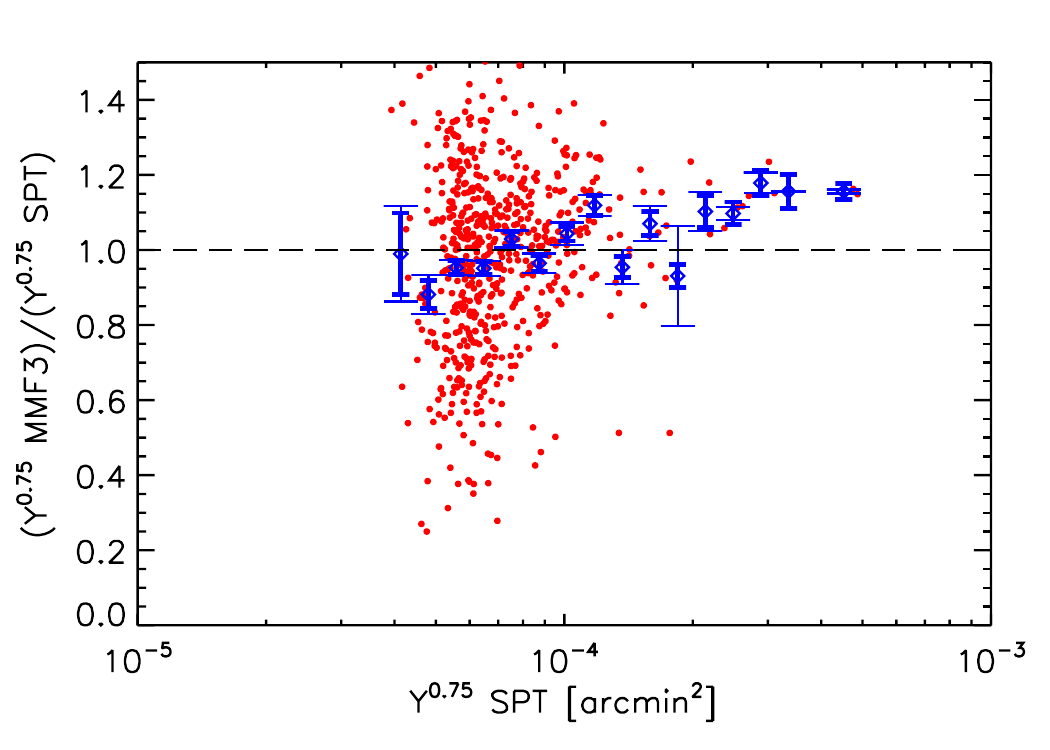}
\caption{{\it Left:} Blind SZ flux of the SPT clusters extracted from the public SPT data with our modified MMF3 algorithm as a function of the SZ flux $Y^{0.75}$ published by the SPT collaboration~\citep{bleem2015}. There is good agreement at low flux, but our flux is systematically overestimated at high values ($Y^{0.75}\, {\rm SPT}>2 \times 10^{-4} \, {\rm arcmin}^2$). {\it Right:} Ratio of the two flux values as a function of the flux published by the SPT collaboration. Our flux is in good agreement for $Y^{0.75}\, {\rm SPT}<2 \times 10^{-4} \, {\rm arcmin}^2$, but it is on average 1.1 times $Y^{0.75}\, {\rm SPT}$ for $Y^{0.75}\, {\rm SPT}>2 \times 10^{-4} \, {\rm arcmin}^2$. Red dots are individual clusters. Blue diamonds are weighted averages. Thick error bars display 68\% statistical errors, and thin error bars show 68\% errors obtained by the bootstrap method.}
              \label{fig:fluxmmf3_vs_fluxspt}
\end{figure*}

Figure~\ref{fig:fluxmmf3_vs_fluxspt} compares our blind SZ flux to the flux published by the SPT collaboration. The left panel shows that there is good agreement between the two flux values at a low flux ($Y^{0.75}\, {\rm SPT}<2 \times 10^{-4} \, {\rm arcmin}^2$), but that our blind flux is systematically overestimated at a high flux ($Y^{0.75}\, {\rm SPT}>2 \times 10^{-4} \, {\rm arcmin}^2$). The right panel shows the ratio of the two flux values: The overestimation is about 10\% for $Y^{0.75}\, {\rm SPT}>2 \times 10^{-4} \, {\rm arcmin}^2$. Since the cluster size is correctly estimated, as shown in Figure~\ref{fig:sizemmf3_vs_sizespt}, we attribute this overestimation to the inadequate modeling of bright sources by the filter transfer function provided by the SPT collaboration, as discussed in Appendix~\ref{sec:filterfunc}.

\section{\Planck\ photometry}
\label{sec:plckphot}

We re-extracted the flux and size of \Planck\ clusters, fixing the position to the coordinates provided in the PSZ2 catalog~\citep{PSZ2} in order to check the consistency between our photometry (i.e., flux and size estimation) and the photometry published by the \Planck\ collaboration. We restricted the PSZ2 cluster catalog to MMF3-only clusters, since our extraction method is derived from it, and we solely focused on the clusters in the SPT footprint.
The differences are not expected to be negligible, because we worked with upgraded \Planck\ maps ($N_{side}=8192$) instead of native \Planck\ maps ($N_{side}=2048$) and we changed the coordinate system from Galactic to equatorial to match the SPT-SZ public data. Although the all-sky maps that we used are the same as the all-sky maps used by the \Planck\ collaboration and thus carry the same information, our $10 \times 10 \deg$ tangential maps around the PSZ2 clusters are different due to the different pixel size and map orientation. This changes the estimation of the noise power spectrum, $ \vec{P}(\vec{k})$. In addition, we did not use the refined point source masking procedure used by the \Planck\ collaboration~\citep[Sect. 3.1 of][]{PSZ2}, but we detected the point sources above $S/N>8$ with single frequency matched filters in individual channel maps and masked them with a ten-arcmin radius disk, which is simpler and computationally faster.

\begin{figure*}
\centering
\includegraphics[width=0.45\hsize]{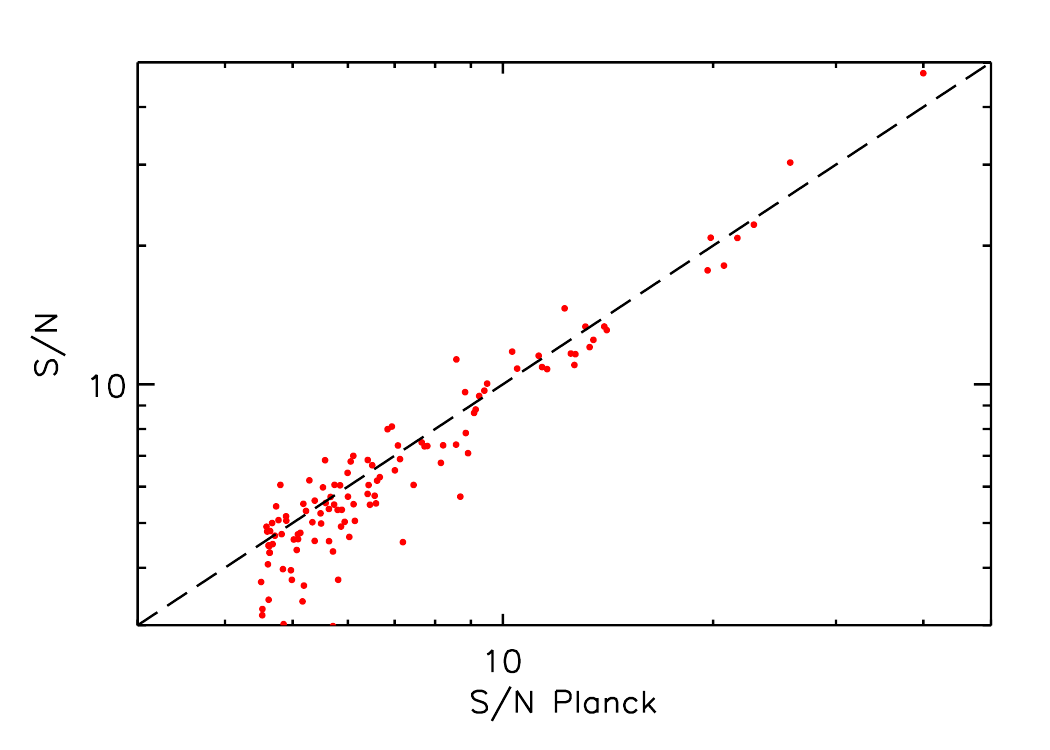}  \includegraphics[width=0.45\hsize]{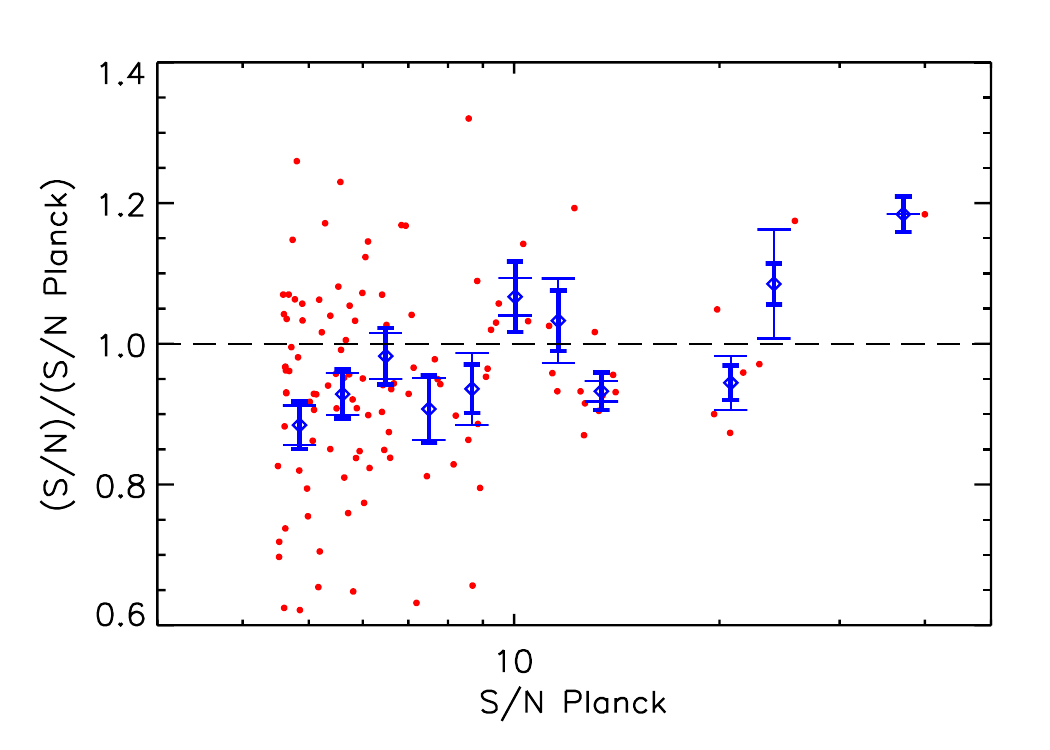}
\caption{{\it Left:} S/N of the \Planck\ clusters extracted from the public \Planck\ data as a function of the signal-to-noise $S/N \, Planck$ published by the \Planck\ collaboration~\citep{PSZ2}. There is overall agreement, but also a large scatter between the two measurements due to the upgraded pixel size, the change of coordinate system, and the change in the point-source masking procedure. {\it Right:} Ratio of the two signal-to-noise values as a function of the signal-to-noise published by the \Planck\ collaboration. Red dots are individual clusters. Blue diamonds are weighted averages. Thick error bars display 68\% statistical errors, and thin error bars show 68\% errors obtained by the bootstrap method.}
              \label{fig:snr_vs_snrplck}
\end{figure*}

\begin{figure*}
\centering
\includegraphics[width=0.45\hsize]{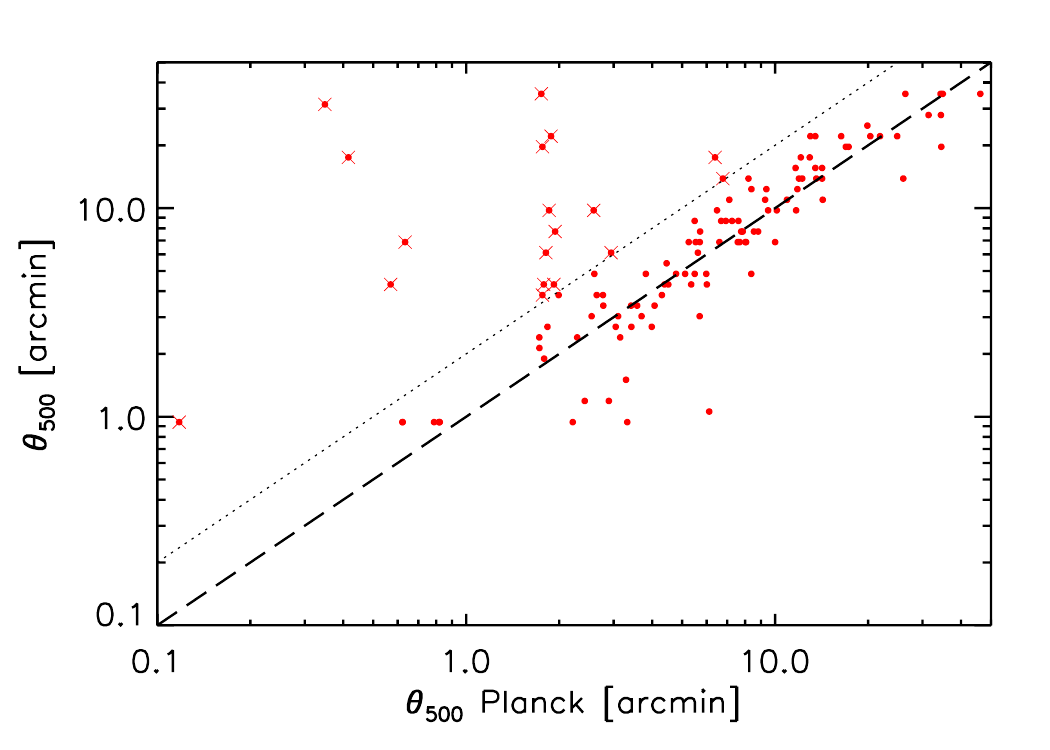}  \includegraphics[width=0.45\hsize]{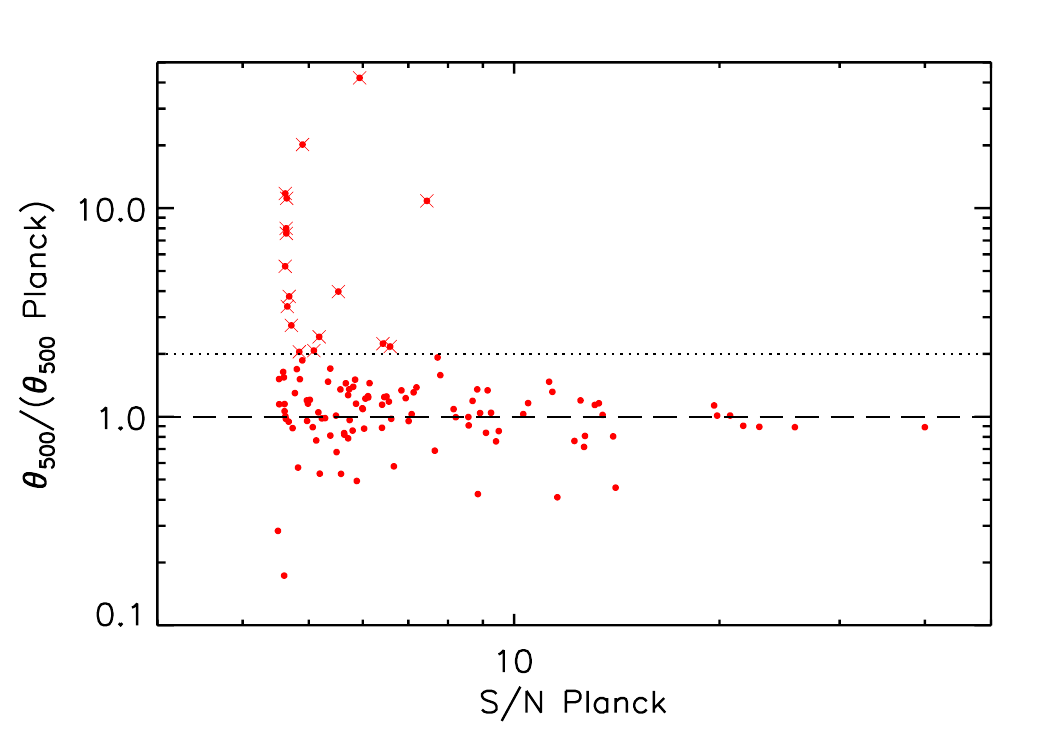}
\caption{{\it Left:} Blind size $\theta_{500}$ of the \Planck\ clusters extracted from the public \Planck\ data as a function of the blind size $\theta_{500} \, Planck$ published by the \Planck\ collaboration~\citep{PSZ2}. Red dots are individual clusters. There is overall agreement, but some clusters show deviations larger than a factor of two. They are marked with a red cross. The dotted black line delineates the deviation by a factor of two. {\it Right:} Ratio of the two size values as a function of the S/N published by the \Planck\ collaboration. The deviating clusters are mainly located at the lower S/N, for which the blind size estimation is uncertain.}
              \label{fig:t500_vs_t500plck}
\end{figure*}

Fig.~\ref{fig:snr_vs_snrplck} shows the results for signal-to-noise. The agreement between our S/N and the S/N published by the \Planck\ collaboration is good, but there is a large scatter due to the upgrades to the pixel size, the change of coordinates, and the difference in the point source masking procedures. \Planck\ blind sizes and fluxes were measured at the location of the maximum of the degeneracy contours provided by the PSZ2 cluster catalog. They were compared to the sizes and fluxes provided by our filter at the maximum of the S/N. The results are presented in Fig.~\ref{fig:t500_vs_t500plck} for the sizes. The blind sizes are also in global good agreement, but some clusters show large deviations as displayed in the left panel. They correspond to clusters detected at low S/N for which the size determination is uncertain, as shown in the right-hand panel. These clusters are marked as red crosses in both panels. 

\begin{figure*}
\centering
\includegraphics[width=0.45\hsize]{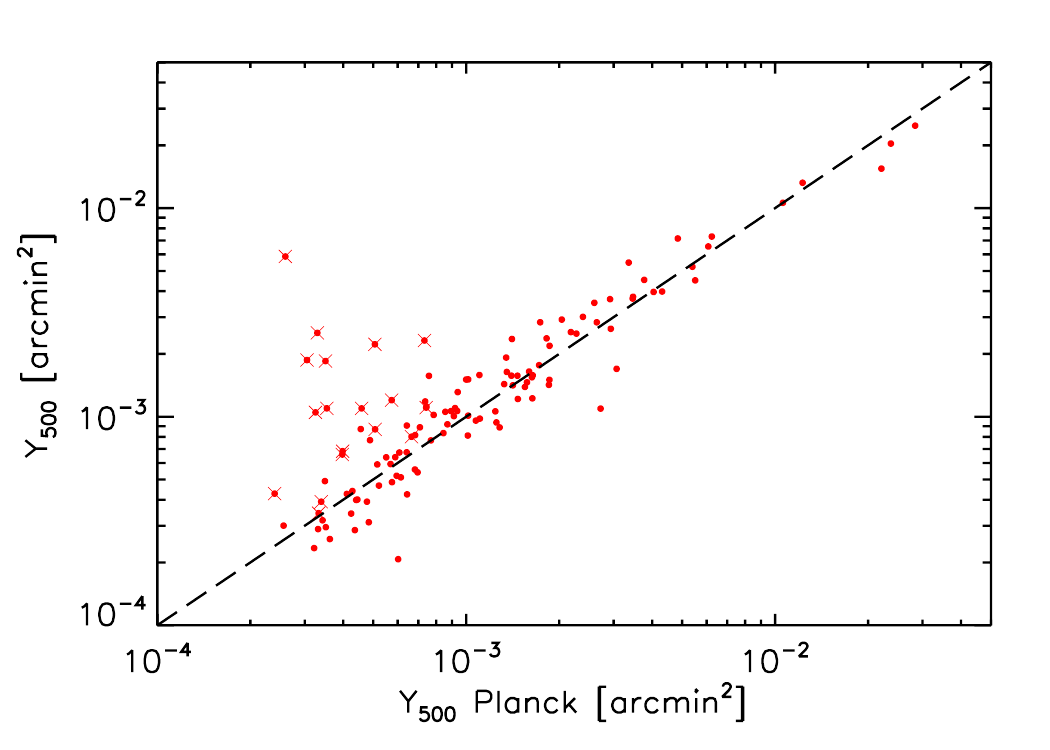}  \includegraphics[width=0.45\hsize]{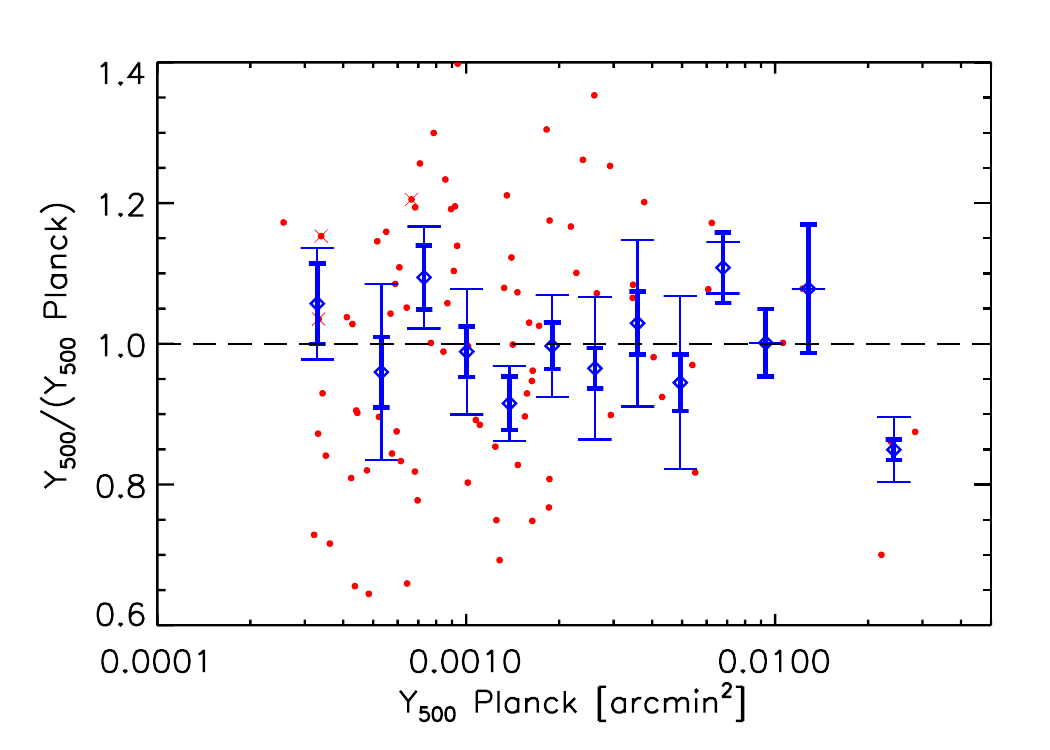}
\caption{{\it Left:} Blind flux $Y_{500}$ of the \Planck\ clusters extracted from the public \Planck\ data as a function of the blind flux $Y_{500} \, Planck$ published by the \Planck\ collaboration~\citep{PSZ2}. Red dots are individual clusters. There is overall agreement, but some clusters show deviations from the equality line. They correspond to clusters having blind sizes deviating by more than a factor of two from the values published by the \Planck\ collaboration and are marked as red crosses (see also Fig.~\ref{fig:t500_vs_t500plck}). {\it Right:} Ratio of the two flux values as a function of the flux published by the \Planck\ collaboration. The ratio is consistent with unity. Blue diamonds are weighted averages. Thick error bars display 68\% statistical errors, and thin error bars show 68\% errors obtained by the bootstrap method.}
              \label{fig:cy500_vs_cy500plck}
\end{figure*}

Fig.~\ref{fig:cy500_vs_cy500plck} shows results for the flux measurements. As for the sizes, the global agreement is good except for some specific clusters. They correspond to clusters that we marked with the red crosses in Fig.~\ref{fig:t500_vs_t500plck}. Due to the low S/N of some detections, the size was poorly determined, which directly translates into a bad recovery of the integrated flux.

\section{Comparison of masses extracted by the joint matched filter and masses published by the SPT and \Planck\ collaborations}
\label{sec:masscomp}

In this appendix, we study outliers in Fig.~\ref{fig:sptsz_match_mass} and \ref{fig:psz2_match_mass} (Sect.~\ref{sec:mass_outliers_sptsz_psz2}) and we investigate the bias in the recovered joint mass with respect to the published mass for PSZ2 clusters (Sect.~\ref{sec:mass_bias_psz2clus}).

\subsection{Outliers in mass}
\label{sec:mass_outliers_sptsz_psz2}

\begin{figure}
\centering
\includegraphics[width=\hsize]{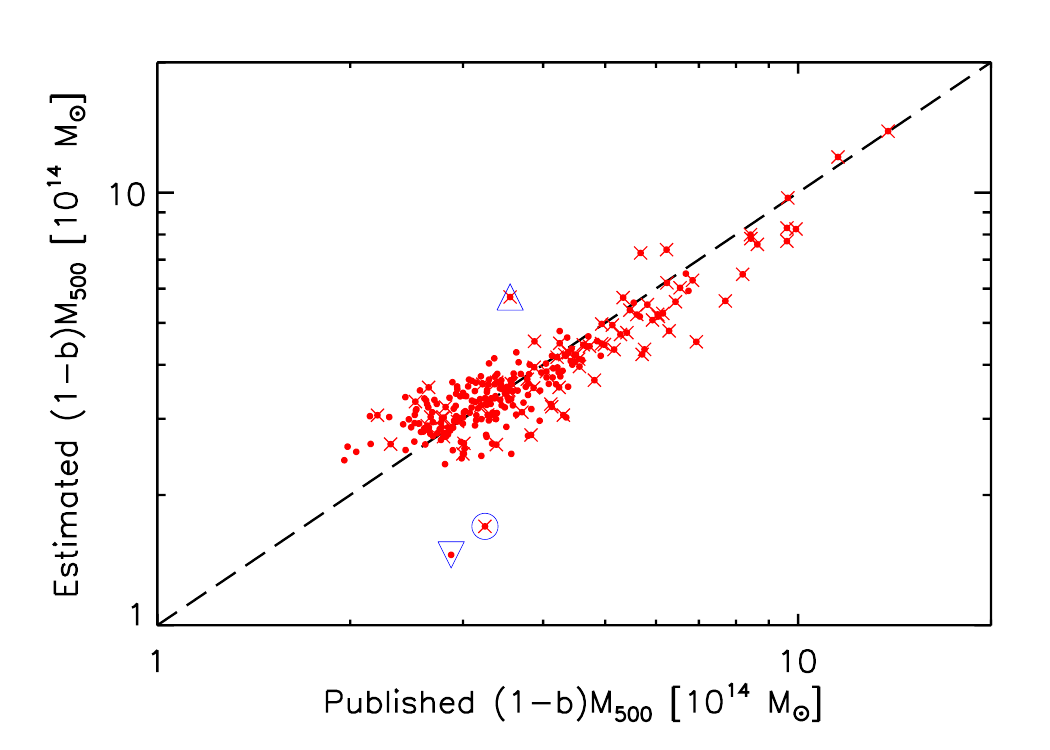}
\caption{Extracted masses versus published SPT-SZ masses for the joint detections matched to SPT-SZ clusters. This figure is the same as Fig.~\ref{fig:sptsz_match_mass}, but we additionally marked the clusters that are both matched to SPT-SZ and PSZ2 clusters with red crosses. We also marked three outliers with a blue circle and upward and downward triangles.}
\label{fig:indepth_sptsz_match_mass}
\end{figure}

\begin{figure}
\centering
\includegraphics[width=\hsize]{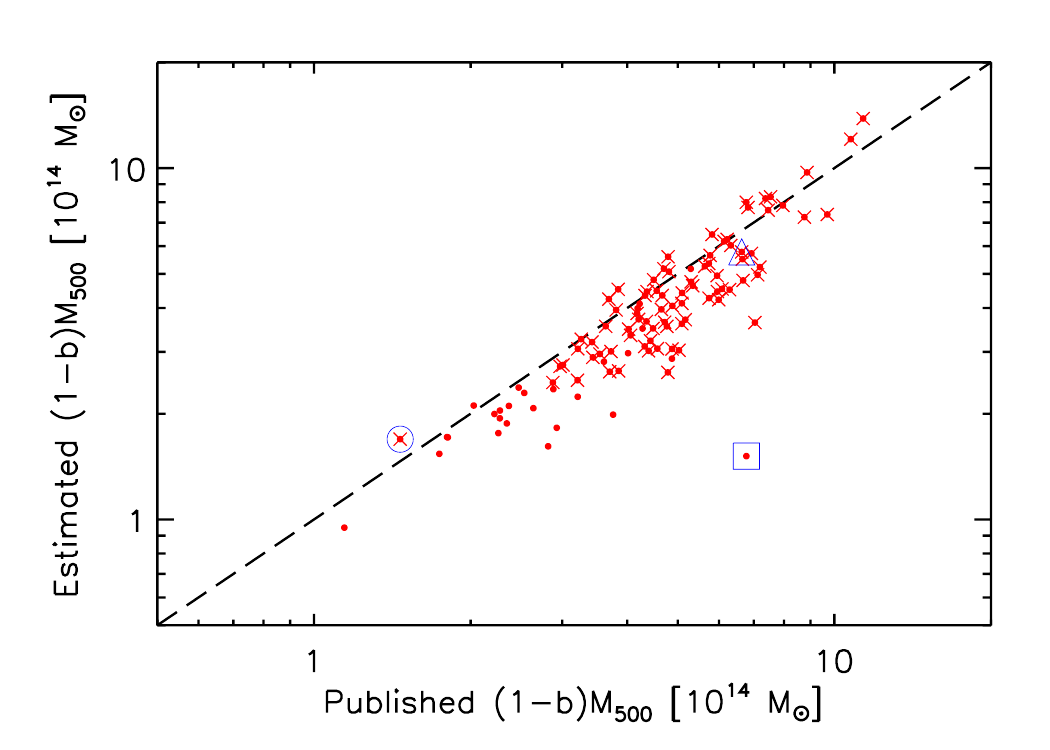}
\caption{Extracted masses versus published PSZ2 masses for the joint detections matched to PSZ2 clusters. This figure is the same as Fig.~\ref{fig:psz2_match_mass}, but we additionally marked the clusters that are both matched to SPT-SZ and PSZ2 clusters with red crosses as in Fig.~\ref{fig:indepth_sptsz_match_mass}. We also marked the main outlier with a blue square. The blue circle and upward triangle, which are outliers in Fig.~\ref{fig:indepth_sptsz_match_mass}, are not outliers in this figure.}
\label{fig:indepth_psz2_match_mass}
\end{figure}

We mark the four clear outliers of Fig.~\ref{fig:sptsz_match_mass} and~\ref{fig:psz2_match_mass} with blue symbols. In Fig.~\ref{fig:indepth_sptsz_match_mass}, the circle marks SPT-CLJ2313-4243 (z=0.0564, also PSZ2 G348.46-64.83), the downward triangle SPT-CLJ0439-5330 (z=0.43), and the upward triangle SPT-CLJ0431-6126 (z=0.0577, also PSZ2 G272.08-40.16). In Fig.~\ref{fig:indepth_psz2_match_mass}, the square marks PSZ2 G271.53-56.57 (z=0.3).

\begin{figure}
\centering
\includegraphics[width=\hsize]{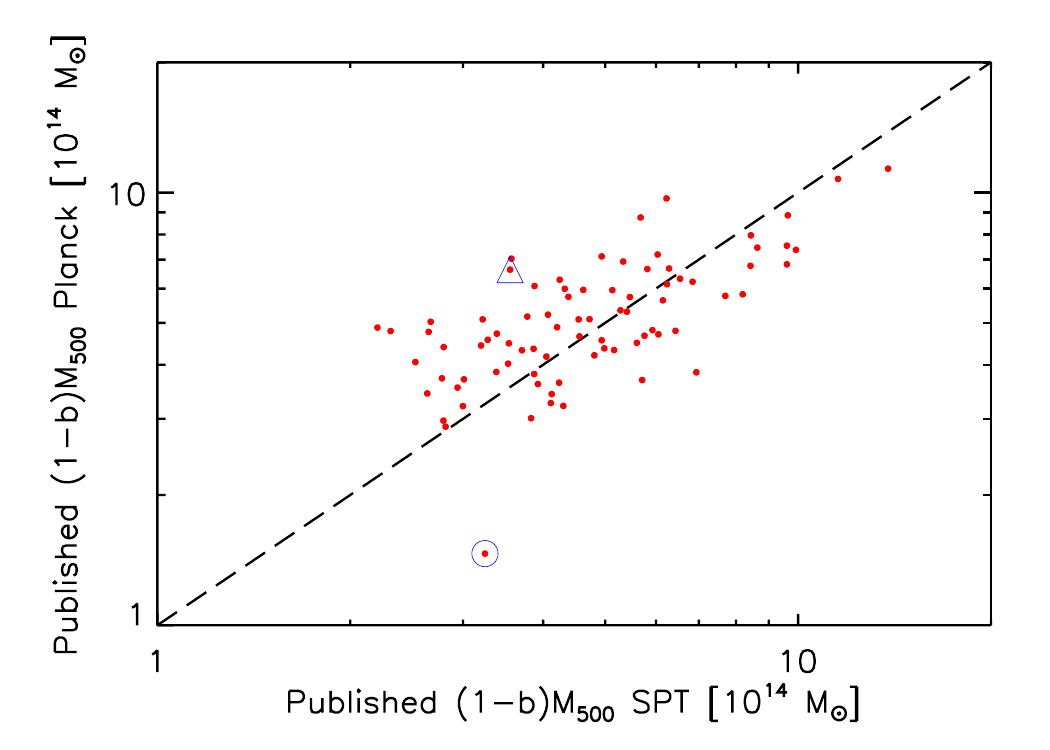}
\caption{Published PSZ2 masses versus published SPT-SZ masses for the clusters that are both in the SPT-SZ and PSZ2 published catalogs. The cluster marked with a blue circle is an outlier.}
\label{fig:mass_spt_psz2}
\end{figure}

We visually inspected the filtered maps of these four clusters and found that SPT-CLJ0431-6126 and PSZ2 G271.53-56.57 are located close to masked point sources, which may have contaminated their fluxes and thus their masses as well. We did not notice any problem in the filtered maps of the two other outliers. SPT-CLJ2313-4243 and SPT-CLJ0431-6126 are marked as outliers in Fig.~\ref{fig:indepth_sptsz_match_mass}, but they are actually not outliers in Fig.~\ref{fig:indepth_psz2_match_mass}. This suggests that the published SPT and \Planck\ masses do not match for these two clusters. Indeed, this is confirmed in Fig.~\ref{fig:mass_spt_psz2}: SPT-CLJ2313-4243 is a clear outlier and SPT-CLJ0431-6126 is at the edge of the distribution. We further investigate the distance between the SPT-SZ and PSZ2 positions for these two clusters as a possible explanation for them being outliers \citep[see Sect.~3.2 of][]{tarrio2019}, but we found the two offsets to be in the overall distribution  of the offsets of the detections matching both SPT-SZ and PSZ2 clusters.

In summary, the outliers may originate from flux contamination due to close by point sources and/or to a miss-match of the masses already present in the published SPT-SZ and PSZ2 catalogs. This mass miss-match in the original catalogs can also be due to point source contamination or specific issues from the datasets (e.g., flux estimation of nearby clusters such as SPT-CLJ2313-4243 and SPT-CLJ0431-6126 in the SPT data may be difficult).
 
\subsection{Joint versus published mass for PSZ2 clusters}
\label{sec:mass_bias_psz2clus}

Fig.~\ref{fig:psz2_match_mass} shows a small underestimation of the joint mass with respect to the published PSZ2 mass. We investigate if this trend is visible for detections matching PSZ2-only or for detections matching both SPT-SZ and PSZ2 clusters. In Fig.~\ref{fig:indepth_psz2_match_mass}, we mark detections matching both SPT-SZ and PSZ2 clusters with red crosses: They show the same trend as detections matching PSZ2-only clusters. We also mark the SPT-SZ and PSZ2 clusters in Fig.~\ref{fig:indepth_sptsz_match_mass} with red crosses: They also show a small underestimation of the joint mass with respect to the published SPT-SZ mass. Detections matching SPT-SZ only do not show a bias.

\begin{figure}
\centering
\includegraphics[width=\hsize]{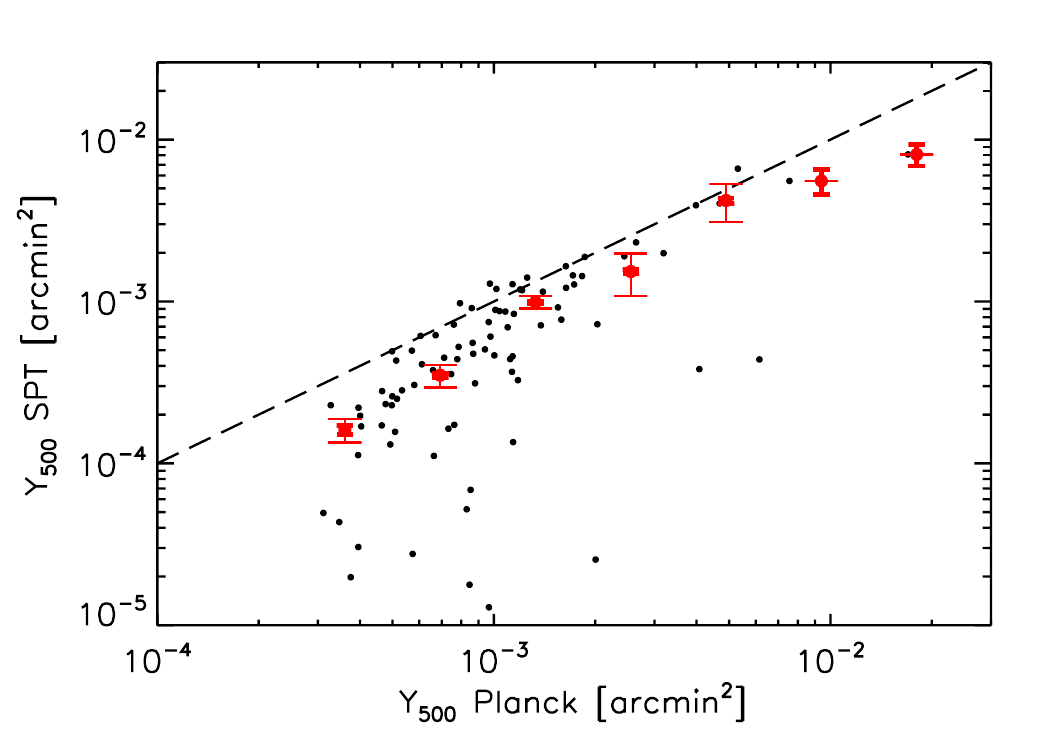}
\caption{SPT fluxes of PSZ2 clusters as a function of PSZ2 flux. This is the same figure as Fig.~\ref{fig:spt_vs_plck_plck}, except that we used the {\em XMM-Newton} prior to fixing the filter size, instead of the blind PSZ2 size, for both the \Planck\ fluxes (x-axis)  and SPT fluxes (y-axis).}
\label{fig:spt_vs_plck_plck_xmm}
\end{figure}

This underestimation of the joint mass with respect to the published mass is very likely due to the underestimation of the SPT flux with respect to \Planck\ flux when using the {\em XMM-Newton} prior to determining the filter size instead of the blind size.
The joint mass is determined using the {\em XMM-Newton} prior. When using this prior to estimate PSZ2 cluster masses in the SPT-SZ data, the SPT flux are systematically smaller than the \Planck\ flux as shown in Fig.~\ref{fig:spt_vs_plck_plck_xmm}.
The consequence of a smaller SPT flux is a smaller joint flux and thus a smaller mass.

In all the figures of Appendix~\ref{sec:masscomp}, the published SPT masses were recalibrated by 0.8 to account for our mass definition.

\section{Names of the missed SPT and PSZ2 clusters}
\label{sec:missed_names}

In Table~\ref{tab:missed_names_spt} and \ref{tab:missed_names_psz2}, we provide the names of the missed SPT and PSZ2 clusters studied in Sect.~\ref{sec:missed}.

\begin{table*}
       \centering
        \caption{The 101 SPT clusters missed in the PSZSPT catalog.}
        \label{tab:missed_names_spt}
         \tiny
        \begin{tabular}{c c c c c c}
SPT-CLJ0001-5440 & SPT-CLJ0002-5557 & SPT-CLJ0011-4614 & SPT-CLJ0013-5714 & SPT-CLJ0015-6000 & SPT-CLJ0021-4902 \\
SPT-CLJ0022-4144 & SPT-CLJ0052-4551 & SPT-CLJ0108-4659 & SPT-CLJ0119-5919 & SPT-CLJ0135-5902 & SPT-CLJ0139-5204 \\
SPT-CLJ0144-4157 & SPT-CLJ0148-4518 & SPT-CLJ0201-6051 & SPT-CLJ0202-5401 & SPT-CLJ0216-5730 & SPT-CLJ0221-4446 \\
SPT-CLJ0231-4427 & SPT-CLJ0242-4150 & SPT-CLJ0257-5732 & SPT-CLJ0257-4817 & SPT-CLJ0313-5334 & SPT-CLJ0337-6300 \\
SPT-CLJ0337-6207 & SPT-CLJ0341-6143 & SPT-CLJ0344-5518 & SPT-CLJ0353-5043 & SPT-CLJ0354-5151 & SPT-CLJ0412-5106 \\
SPT-CLJ0418-4552 & SPT-CLJ0421-4845 & SPT-CLJ0422-4608 & SPT-CLJ0428-6049 & SPT-CLJ0430-6251 & SPT-CLJ0433-5630 \\
SPT-CLJ0444-5603 & SPT-CLJ0444-4352 & SPT-CLJ0454-4211 & SPT-CLJ0455-4159 & SPT-CLJ0456-4906 & SPT-CLJ0459-4947 \\
SPT-CLJ0500-5116 & SPT-CLJ0512-5139 & SPT-CLJ0516-5755 & SPT-CLJ0522-5026 & SPT-CLJ0528-4417 & SPT-CLJ0529-6051 \\
SPT-CLJ0533-5005 & SPT-CLJ0535-5956 & SPT-CLJ0539-6013 & SPT-CLJ0540-5744 & SPT-CLJ0550-6358 & SPT-CLJ0559-6022 \\
SPT-CLJ0611-4724 & SPT-CLJ0617-5507 & SPT-CLJ0622-4645 & SPT-CLJ0625-4330 & SPT-CLJ0637-6112 & SPT-CLJ0637-4327 \\
SPT-CLJ0643-4535 & SPT-CLJ0643-5056 & SPT-CLJ0649-4510 & SPT-CLJ0650-4503 & SPT-CLJ2011-5725 & SPT-CLJ2020-6314 \\
SPT-CLJ2026-4513 & SPT-CLJ2030-5638 & SPT-CLJ2040-5342 & SPT-CLJ2040-4451 & SPT-CLJ2043-5035 & SPT-CLJ2051-6256 \\
SPT-CLJ2056-5459 & SPT-CLJ2058-5608 & SPT-CLJ2101-6123 & SPT-CLJ2106-4421 & SPT-CLJ2108-4445 & SPT-CLJ2109-5040 \\
SPT-CLJ2111-5339 & SPT-CLJ2120-4728 & SPT-CLJ2136-4704 & SPT-CLJ2140-5727 & SPT-CLJ2143-5509 & SPT-CLJ2145-4348 \\
SPT-CLJ2152-5633 & SPT-CLJ2155-5224 & SPT-CLJ2155-6048 & SPT-CLJ2158-5451 & SPT-CLJ2203-5047 & SPT-CLJ2205-5927 \\
SPT-CLJ2232-6151 & SPT-CLJ2235-4416 & SPT-CLJ2254-5805 & SPT-CLJ2259-5431 & SPT-CLJ2301-5546 & SPT-CLJ2308-4834 \\
SPT-CLJ2311-4203 & SPT-CLJ2311-5820 & SPT-CLJ2321-5419 & SPT-CLJ2352-5846 & SPT-CLJ2353-5512 &    \\
        \end{tabular}
\end{table*}

\begin{table*}
       \centering
        \caption{The 11 PSZ2 clusters missed in the PSZSPT catalog.}
        \label{tab:missed_names_psz2}
         \tiny
        \begin{tabular}{c c c c c c}
        PSZ2 G265.21-24.83 & PSZ2 G341.09-33.15 & PSZ2 G331.96-45.74 & PSZ2 G255.52-35.66 & PSZ2 G327.66-54.26 & PSZ2 G336.01-51.27 \\
PSZ2 G269.36-47.20 & PSZ2 G319.64-65.11 & PSZ2 G265.60-46.87 & PSZ2 G299.68-60.21 & PSZ2 G279.51-44.85 &    \\
        \end{tabular}
\end{table*}

\section{Description of the PSZSPT catalog}
\label{sec:description}

The PSZSPT catalog contains 419 detections. For each detection, we provide the following fields, partially shown in Table~\ref{tab:format_cat}. A complete version of the catalog is provided in electronic format.
\begin{itemize}
\item NAME: Name of the candidate, PSZSPT Jxxxx+yyyy
\item RAJ2000: Right ascension (J2000) in degrees
\item DEJ2000: Declination (J2000) in degrees
\item GLON: Galactic longitude in degrees
\item GLAT: Galactic latitude in degrees
\item SNR: Signal-to-noise ratio obtained with the best filter size
\item RANK: Rank of the candidate (0=unidentified; 1=identified; 2=possibly identified; 3=multiple detection)
\item Z: Redshift of the candidate
\item Z\_REF: Origin of the redshift
\item M500: Estimated cluster mass in solar masses
\item M500\_INF: Lower bound of 68\% confidence interval on the estimated cluster mass in solar masses
\item M500\_SUP: Upper bound of 68\% confidence interval on the estimated cluster mass in solar masses
\item SPT: Name of the SPT \citep{bleem2015} cluster matched to the candidate
\item PSZ2: Name of the \Planck\ PSZ2 \citep{PSZ2} cluster matched to the candidate
\item MCSZ: Name of the MCSZ (\url{https://www.galaxyclusterdb.eu/}) cluster matched to the candidate
\item MCXC: Name of the MCXC \citep{piffaretti2011} cluster matched to the candidate
\item COMPRASS: Name of the ComPRASS \citep{tarrio2019} cluster matched to the candidate
\item ABELL: Name of the Abell \citep{abell1989} cluster matched to the candidate
\item BCS: Name of the BCS \citep{desai2012, bleem_blanco2015} cluster matched to the candidate
\item MARD: Name of the MARD-Y3 \citep{klein2019} cluster matched to the candidate
\item WHY: Name of the WHY \citep{wen2018} cluster matched to the candidate
\item SIMBAD: Name of the SIMBAD counterpart found in a 20 arcmin radius disk around the candidate
\item NED: Name of the NED counterpart found in a 20 arcmin radius disk around the candidate
\item NOTES: Notes on specific candidates
\end{itemize}

%\begin{sidewaystable*}
\begin{table*}
       %\centering
        \caption{Cluster candidates in the PSZSPT catalog. The different fields are described in the text of this appendix. This table is available in its entirety in a machine-readable format.}
        \label{tab:format_cat}
        \tiny
        \begin{tabular}{c c c c c c c c c c c c}
                \hline
                \noalign{\smallskip}
                NAME & RAJ2000  & DEJ2000 &  GLON   & GLAT   &   S/N & RANK & Z & Z\_REF & M500 & M500\_INF & M500\_SUP \\
                            &      [deg]            &        [deg]           &         [deg]          &       [deg]         &           &             &     &            & [$10^{14} M_\odot$] & [$10^{14} M_\odot$] & [$10^{14} M_\odot$] \\
                \noalign{\smallskip}
                \hline
                \noalign{\smallskip}
PSZSPT J0000-4356 & 0.061 & -43.949 & 331.122 & -70.270 & 5.19 & 1 & 1.000 & SPT & 2.61 & 2.23 & 2.97 \\
PSZSPT J0000-5748 & 0.241 & -57.808 & 315.638 & -58.056 & 7.40 & 1 & 0.701 & SPT & 3.48 & 3.15 & 3.80 \\
PSZSPT J0001-4843 & 0.292 & -48.718 & 323.868 & -66.315 & 6.65 & 1 & 0.300 & SPT & 3.17 & 2.81 & 3.51 \\
PSZSPT J0003-5253 & 0.840 & -52.889 & 318.846 & -62.727 & 5.07 & 1 & 0.279 & WHY & 2.50 & 2.13 & 2.84 \\
PSZSPT J0012-5352 & 3.073 & -53.872 & 315.381 & -62.299 & 6.15 & 1 & 0.330 & SPT & 2.82 & 2.47 & 3.15 \\
PSZSPT J0013-4907 & 3.320 & -49.117 & 318.920 & -66.808 & 12.3 & 1 & 0.406 & SPT & 5.17 & 4.86 & 5.47 \\
PSZSPT J0014-4952 & 3.694 & -49.876 & 317.661 & -66.190 & 11.1 & 1 & 0.752 & SPT & 4.78 & 4.48 & 5.07 \\
PSZSPT J0015-5303 & 3.861 & -53.064 & 314.969 & -63.213 & 5.10 & 0 & -1.00 &  & -1.0 & -1.0 & -1.0 \\
PSZSPT J0019-5527 & 4.819 & -55.452 & 312.372 & -61.081 & 7.02 & 1 & 0.755 & SPT & 3.39 & 3.06 & 3.71 \\
PSZSPT J0025-4133 & 6.490 & -41.551 & 321.219 & -74.659 & 6.39 & 1 & 0.430 & SPT & 2.87 & 2.57 & 3.16 \\
... & & & & & & & & & & &
        \end{tabular}

                \begin{tabular}{c c c c c}
                \noalign{\smallskip}
                \noalign{\smallskip}
                \hline
                \noalign{\smallskip}
                NAME & SPT  & PSZ2 & MCSZ & MCXC \\
                \noalign{\smallskip}
                \hline
                \noalign{\smallskip}
PSZSPT J0000-4356 & SPT-CLJ0000-4356 &  & SPT-CLJ0000-4356 &  \\
PSZSPT J0000-5748 & SPT-CLJ0000-5748 &  & SPT-CLJ0000-5748 &  \\
PSZSPT J0001-4843 & SPT-CLJ0001-4842 &  & SPT-CLJ0001-4842 &  \\
PSZSPT J0003-5253 &  &  &  &  \\
PSZSPT J0012-5352 & SPT-CLJ0012-5352 &  & SPT-CLJ0012-5352 &  \\
PSZSPT J0013-4907 & SPT-CLJ0013-4906 &  & SPT-CLJ0013-4906 &  \\
PSZSPT J0014-4952 & SPT-CLJ0014-4952 &  & SPT-CLJ0014-4952 &  \\
PSZSPT J0015-5303 &  &  &  &  \\
PSZSPT J0019-5527 & SPT-CLJ0019-5527 &  & SPT-CLJ0019-5527 &  \\
PSZSPT J0025-4133 & SPT-CLJ0025-4133 &  & SPT-CLJ0025-4133 &  \\
... & & & & 
        \end{tabular}

               \begin{tabular}{c c c c c c}
                \noalign{\smallskip}
                \noalign{\smallskip}
                \hline
                \noalign{\smallskip}
                NAME & COMPRASS & ABELL & BCS & MARD & WHY \\
                \noalign{\smallskip}
                \hline
                \noalign{\smallskip}
PSZSPT J0000-4356 &  & S1173 &  &  & WHY J000124.5-435522 \\
PSZSPT J0000-5748 &  &  &  &  &  \\
PSZSPT J0001-4843 &  &  &  &  & WHY J000004.3-483442 \\
PSZSPT J0003-5253 &  &  &  &  & J000314.6-525516 \\
PSZSPT J0012-5352 &  &  &  &  & J001216.7-535103 \\
PSZSPT J0013-4907 & PSZRX G318.95-66.81 &  &  & MARDJ001314.2-490541 & WHY J001418.2-491852 \\
PSZSPT J0014-4952 &  & 2753 &  &  & J001626.6-495035 \\
PSZSPT J0015-5303 &  &  &  &  &  \\
PSZSPT J0019-5527 &  &  &  &  &  \\
PSZSPT J0025-4133 &  &  &  &  &  \\
... & & & & &
        \end{tabular}

                \begin{tabular}{c c c c}
                \noalign{\smallskip}
                \noalign{\smallskip}
                \hline
                \noalign{\smallskip}
                NAME & SIMBAD & NED & NOTES \\
                \noalign{\smallskip}
                \hline
                \noalign{\smallskip}
PSZSPT J0000-4356 &  &  &  \\
PSZSPT J0000-5748 &  &  &  \\
PSZSPT J0001-4843 &  &  &  \\
PSZSPT J0003-5253 &  &  &  \\
PSZSPT J0012-5352 &  &  &  \\
PSZSPT J0013-4907 &  &  &  \\
PSZSPT J0014-4952 &  &  &  \\
PSZSPT J0015-5303 & SPT-CL J0013-5310 & SPT-CL J0013-5310 & SPT-CL J0013-5310 @ 17.4 arcmin is not associated to another PSZSPT detection \\
PSZSPT J0019-5527 &  &  &  \\
PSZSPT J0025-4133 &  &  &  \\
... & & &
        \end{tabular}

%\end{sidewaystable*}
\end{table*}

\bibliographystyle{aa}
\bibliography{pszspt}

\end{document}